\documentclass[12pt]{article}

\usepackage{hyperref}
\pdfoutput=1
\usepackage{graphics}
\usepackage[english] {babel}
\usepackage{amssymb,amsfonts,amsmath,mathtext}
\usepackage{graphicx}
\usepackage{xcolor}

\usepackage{subfig}
\usepackage{floatrow}

\usepackage{float}
\floatstyle{plaintop}
\restylefloat{table}

\usepackage{natbib}
\DeclareMathOperator{\Ex}{\mathbb{E}} 
\newcommand{\Var}{{\mathsf{Var}}\,}

\renewcommand{\S}{{\mathbb S}}

\DeclareMathOperator{\sd}{SD}

\renewcommand{\d}{{\rm d}}

\newcommand{\etc}{etc.\ }
\newcommand{\eg}{e.g.,\ }

\newcommand{\vs}{vs.\ }
\newcommand{\ie}{i.e., }

\newcommand{\e}{\mathrm{e}}
\renewcommand{\i}{\mathsf{i}}
   \title{Impact of non-stationarity on hybrid ensemble filters: A study with a
   doubly stochastic advection-diffusion-decay model\footnote{
        This is the accepted version of the article published in final form in
        {\em Quart. J. Roy. Meteorol. Soc.}, 2019, v. 145, N 722, 2255--2271,  DOI: 10.1002/QJ.3556.
        }
        }
   \author{Michael Tsyrulnikov and Alexander Rakitko\\
   \smallskip
   {\small (michael.tsyrulnikov@gmail.com) }}

\begin{document}

\maketitle
\floatsetup[figure]{style=plain,subcapbesideposition=top}

\begin{abstract}

Effects of non-stationarity on the performance of hybrid ensemble filters are studied
(by hybrid filters we mean those which blend ensemble covariances with some other
regularizing  covariances).
To isolate effects of non-stationarity from effects  due to nonlinearity (and the
non-Gaussianity it causes), 
a new doubly stochastic advection-diffusion-decay model (DSADM) is proposed.
The model is hierarchical: it is a linear stochastic
partial differential equation whose coefficients are 
random fields defined through their own stochastic partial differential equations.
DSADM generates conditionally Gaussian spatiotemporal random fields with  
a tunable degree of non-stationarity in space and time.
DSADM allows the use of the exact Kalman filter as a baseline benchmark.

In numerical experiments with DSADM as the ``model of truth'', 
the relative importance of the three kinds of covariance blending is studied:  with 
static, time-smoothed, and space-smoothed covariances.
It is shown that the stronger the non-stationarity, the less useful the static covariance matrix becomes
and the more beneficial the time-smoothed covariances are.
Time-smoothing of background-error covariances proved to be systematically  
more useful than their space-smoothing.
Under non-stationarity, a filter that 
extends the (previously proposed by the authors) Hierarchical Bayes Ensemble Filter
and accommodates the three covariance-blending techniques is shown to outperform
all other configurations of the filters tested.
The R code of the model and the filters is available from github.com/cyrulnic/NoStRa.
\end{abstract}


{\bf Running head}: Impact of non-stationarity on hybrid  filters

\section {Introduction}

Ensemble Kalman filters (EnKF) estimate background-error covariances from an ensemble
of forecasts. The resulting sample covariances are noisy and rank deficient if the ensemble size
is not greater than the dimensionality of the covariance matrix (which is typically the case in 
geosciences). A kind of regularization (that is, introduction of additional information
to stabilize the solution) is needed to make 
them a useful estimate of  true background-error covariances.
Two regularization techniques are most popular: localization and hybridization.
Our focus in this study is on the second technique, in which sample covariances are 
blended (mixed) with some regularizing covariances and which we therefore call ``covariance blending''.

From the literature, we can identify three main types of covariance blending.
First, hybrid ensemble-variational filters (EnVar) employ blending with static ``climatological'' covariances
\citep[][]{HamillSnyderEV,LorencEV,BuehnerEV}.
In statistical literature this regularization device is known as a shrinkage estimator \citep{Ledoit}.
Second,  in the Hierarchical Bayes Ensemble Filter (HBEF) by \citet[][]{TsyRa},
ensemble covariances are blended with time-smoothed recent-past covariances.
The idea of accommodating recent-past background-error statistics has also  been explored by other authors:
\citet[][]{Gustafsson} use time-lagged ensemble members,
\citet[][]{Berre}  use ensemble members from past 4 days to increase the ensemble size,
\citet[][]{Bonavita} use ensemble covariances from previous 12 days to estimate their 
parametric covariance model, 
\citet[][]{Lorenc} found that using time-lagged and time-shifted perturbations
increases the effective ensemble size.
Third, \citet{BuehnerCharron,MenetrierBerre,MenetrierAuligne} studied 
spatial smoothing  of ensemble covariances, which also can  be considered as covariance blending,
this time with neighboring covariances in space.

With any of the above  kinds of covariance blending, 
the intention is to reduce  sampling noise and increase the
rank of the covariance matrix --- but at the expense of reducing flow dependence 
(the main advantage of ensemble statistics)
in the blended covariances.
The question arises: what are the optimal weights of the regularizing covariances in the blend
and which factors  do they depend on?

First of all, we notice that if a data assimilation system is stationary, that is, 
its true (actual) spatial error covariances do not change in time,
then there is no need for an  (on-line) ensemble at all. 
Indeed, with  constant-in-time background-error covariances, all one has to do is to 
carefully estimate (off-line)  their spatial covariance matrix 
(through temporal averaging of innovation-based  or ensemble covariances).
In the Gaussian case, the estimated static covariance matrix will be optimal 
\citep[see, e.g., Theorem 2.38 in][for a formal statement]{Kumar}.
Thus, we conclude that only under {\em non-stationarity} ensemble filters can outperform 
data assimilation schemes that employ static background-error
covariances. True background-error covariances 
in a data assimilation scheme are non-stationary either if the system is governed 
by a non-stationary model or the observation network is changing in time (or both, of course).
In this research, our focus is on non-stationarity caused by the  model.
To study the impact of non-stationarity  on the optimal design of hybrid filters 
we seek a toy model that is capable of producing spatiotemporal fields with 
tunable non-stationarity.

Most toy  models used in data assimilation studies  are nonlinear.
The simplest  models have just a few variables, like
the Ikeda model \citep[][]{Hansen}, 
the double well model \citep[][]{Miller}, and the most popular 
three-variable  Lorenz'63 model \citep[][]{lorenz63}.
Models  defined on a 1D spatial domain   include, among others,
the viscous Burgers equation \citep[][]{Apte},
the Korteweg--de Vries equation  \citep[][]{Bennett},
and the most popular Lorenz'96 and Lorenz'2005 models \citep[][]{Lorenz1998,Lorenz2005}. 
In nonlinear models, the tangent-linear model operator, generally, varies in time, giving rise 
to non-stationarity  (flow-dependence) of filtering probability distributions.
However, with nonlinear models, non-stationarity is inevitably
intertwined with nonlinearity of the evolution of filtering errors and their ensuing non-Gaussianity. 
To disentangle effects due to nonlinearity
from effects of non-stationarity,
we propose  a model that gives rise to non-stationary filtering distributions 
while  being linear and Gaussian.

We build the model on the time-discrete {\em scalar} (\ie one-variable) model
introduced by  \citet{TsyRa}:
\begin {equation}
\label{emdlx}
\xi_{k}=F_{k} \xi_{k-1} + \sigma_{k} \alpha_k,
\end {equation}
where $\xi_{k}$ is the (scalar) true system state,  $k$ labels the time instant,
$F_{k}$ is the (scalar) model operator,  $\sigma_{k}$ is the standard deviation of the 
forcing, and $\alpha_k  \sim {\cal N}(0,1)$ is the driving white noise.
The model is  {\em doubly stochastic} \citep[e.g.][]{tjostheim}, which
means that not only the forcing $\alpha_k$ is random,
the coefficients $F_{k}$  and ${\sigma}_{k}$ are 
random sequences by themselves, each defined through its own
stochastic model similar to Eq.(\ref{emdlx}) but with a constant model operator and 
constant magnitude of the forcing.
{\em Conditionally on the processes} $F_{k}$  and ${\sigma}_{k}$
(that is, after their realizations are computed and kept fixed), 
Eq.(\ref{emdlx}) is a linear model with time-varying coefficients
\citep[e.g.][Section 11.2.1]{Chatfield},
whose solution is a non-stationary random process.
In this study we extend the model  Eq.(\ref{emdlx}) to
the spatiotemporal context.
The result is a new  doubly stochastic advection-diffusion-decay model (DSADM),
whose solutions are non-stationary conditionally Gaussian random fields.

The general idea of non-stationary random field modeling
by assuming that parameters of a spatial or spatiotemporal model are random fields themselves
is discussed by several authors.
\citet[][]{Piterbarg} study an advection-diffusion model whose coefficients are random fields
with prescribed covariance functions.
\citet[][]{Lindgren} allow a length-scale parameter and a variance parameter of their
stochastic partial differential equation to slowly vary in space (by expanding them in a set of
 basis functions).
\citet[][sec. 11.6]{Banerjee} mention the possibility of formulating a
stochastic  differential equation for parameters of another stochastic differential equation.
\citet{Roininen2016} use a stochastic elliptic equation to  model a spatial random field 
and introduce a  hypermodel for its local length scale to achieve non-stationarity;
the hypermodel employs the same  stochastic elliptic equation.
\citet{Dunlop} discuss (among others) more general 
Gaussian hierarchical processes, which are defined through a hierarchy of levels
so that at each level, the conditional probability distribution is Gaussian given 
the previous (parent) level.
Our innovation is the non-stationary hierarchical model
with spatiotemporal stochastic partial differential equations at two levels in the hierarchy.
The particular pattern of the non-stationarity (that is, how the field's variance, local space
and time scales, \etc vary in space-time) is random, and its spatiotemporal structure is 
highly tunable by the model's hyperparameters.

The rest of the paper is organized as follows.
In sections \ref{sec_hom}--\ref{sec_behav} we motivate, describe, and examine 
in numerical experiments the new DSADM model.
In section \ref{sec_hhbef} we introduce a new hybrid-HBEF (named HHBEF) filter,
which accommodates the three covariance-blending techniques mentioned above.
In section \ref{sec_perfHybr} we present results of numerical experiments
with hybrid ensemble filters showing the crucial impact of non-stationarity
on the filters' performance and optimal design.
Efficient numerical implementation of hybrid filters 
is beyond the scope of this study.

The R code of the doubly stochastic model,  the filters, 
and the R scripts that produced this paper's  numerical results 
are  available from https://github.com/cyrulnic/NoStRa.
More detailed texts on the DSADM's numerical scheme, generation of initial
conditions, specification of model parameters, \etc 
can also be found there.

\section {Stochastic advection-diffusion-decay model with constant coefficients}
\label{sec_hom}

To introduce notation and motivate the new doubly stochastic advection-diffusion-decay model,
we theoretically examine here  statistics of  solutions to a {\em stationary} 
stochastic advection-diffusion-decay model with non-stochastic
and constant coefficients (parameters). 
Specifically, we identify how the model parameters affect the variance, 
the spatial scale, and the time scale of the solution.
This is a preparatory section with largely background material.

\subsection {Model}

The model is the following stochastic partial differential equation
(\citet[][Ch. 20, Sec. 3]{Whittle}, \citet[][Eq.(17)]{Lindgren},
 \citet[][]{Sigrist}):
\begin {equation}
\label{mdl}
\frac{\partial\xi} {\partial t} + U \frac{\partial\xi} {\partial s} + \rho\xi -
    \nu \frac{\partial^2\xi} {\partial s^2} = \sigma \alpha,
\end {equation}
where $t$ is time, $s$ is the spatial coordinate on the circle $\S^1(R)$ of radius $R$,
 $U$ is the  advection velocity,
 $\rho$ is the decay (damping) parameter, 
 $\nu$ is the  diffusion parameter, 
  $\alpha(t,s)$ is the standard white (in space and time) Gaussian noise, and
  $\sigma$ is the intensity of the forcing.
The four parameters $\boldsymbol\theta=(U,\rho,\nu,\sigma)$ are constant in space and time.


\subsection {Stationary spatiotemporal statistics}

We start with rewriting Eq.(\ref{mdl}) using the material time derivative (\ie
along the characteristic $s =s_0 +  U t$), or, equivalently, 
switching to the Lagrangian frame of reference by  making the change of variables
$(t,s) \mapsto (t, s- U t)$:
\begin {equation}
\label{ad2}
\frac{\d\xi} {\d t}  +\rho\xi - \nu \frac{\partial^2\xi} {\partial s^2} = \sigma\alpha.
\end {equation}
Next, we employ the spectral expansion in space (retaining only those spectral components
that are resolved on the regular spatial grid with $n$ points),
\begin {equation}
\label{xispe}
\xi(t,s) = \sum_{m=-n/2+1}^{n/2} \tilde\xi_m(t) \,\e^{\i {m s}/{R}}
\end {equation}
and
\begin {equation}
\label{aspe}
\alpha(t,s) = \sum_{m=-n/2+1}^{n/2} \tilde\alpha_m(t) \, \e^{\i {m s}/{R}},
\end {equation}
where  $\i$ is the imaginary unit
and $ \tilde\xi_m(t)$ and $\tilde\alpha_m(t)$ are the (complex) spectral coefficients.
It can be shown  \citep[e.g.][Appendix A.4]{TsyGay_ar} that  $\tilde\alpha_m(t)$ are 
independent standard complex-white-noise
processes $\omega_m(t)$ with  the common intensity  $a={1} /{\sqrt{2\pi R}}$:
\begin {equation}
\label{alpham}
\tilde\alpha_m(t) = a\, \omega_m(t).
\end {equation}
Now, we substitute Eqs.(\ref{xispe})--(\ref{alpham}) into Eq.(\ref{ad2}), getting
\begin {equation}
\label{mdl_spe}
\frac{\d\tilde\xi_m} {\d t} + (\rho + \frac{\nu} {R^2} m^2) \tilde\xi_m = a\sigma \omega_m(t).
\end {equation}
This is the spectral-space form of the model Eq.(\ref{ad2}).
It is easily seen that if $\rho + \frac{\nu} {R^2} m^2>0$,
the solutions to Eq.(\ref{mdl_spe}) for different $m$ become, after an initial transient, 
mutually independent stationary zero-mean random processes\footnote{
Indeed, the influence of the initial condition on $\tilde\xi_m(t)$ exponentially
decays in time, leaving $\tilde\xi_m(t)$ dependent only on the  noise process $\omega_m(t')$ for 
$0\le t' \le t$. The mutual independence of $\tilde\xi_m(t)$ for different $m$
then follows from the mutual independence of the driving noises $\omega_m(t)$.
}.

This implies that the physical-space solution $\xi(t,s)$ becomes 
a zero-mean random field that is stationary in time and space.
Note that by definition, the zero-mean real-valued random field (process) $\xi(t,s)$  
is (second-order) stationary 
if its spatiotemporal covariance function 
$\gamma(t_1,s_1;\, t_2,s_2)= \Ex \xi(t_1,s_1)\, {\xi(t_2, s_2)}$ 
is invariant under translations:
$\Ex \xi(t_1,s_1)\,\xi(t_2,s_2) = \Ex \xi(t_1+u,s_1+v)\,\xi(t_2+u,s_2+v)$
and thus is a function of the space and time shifts only:
$\gamma(t,s)= \Ex \xi(t_1,s_1)\, {\xi(t_1+t, s_1+s)}$.
Here periodicity in the spatial coordinate
$s$ is of course assumed, $\xi(t, s) = \xi(t, s+2\pi R)$.

Each elementary stochastic process $\tilde\xi_m(t)$ is an Ornstein-Uhlenbeck process \citep[e.g.][sec. 8.3]{Arnold}
with the stationary covariance function
\begin {equation}
\label{Bmt}
B_m(t) = b_m \cdot\e^{ -{|\mathnormal{t}|}/\mathnormal{\tau_m}  },
\end {equation}
where the spectral variances $b_m$ are 
\begin {equation}
\label{bm}
 b_m = \frac{a^2 \sigma^2}{2} \cdot \frac {1} {\rho + \frac {\nu} {R^2} m^2}
\end {equation}
and the  spectral time  scales $\tau_m$ are
\begin {equation}
\label{tau_spe}
  \tau_m = \frac{1}  {\rho + \frac {\nu} {R^2} m^2}. 
\end {equation}
Note that Eq.(\ref{tau_spe}) provides the motivation for the inclusion of the decay term in the model.
Indeed, with $\rho=0$, the time scale $\tau_0$  would be infinitely large,
which, on the finite sphere, is unphysical (specifying $b_0=0$ could resolve the problem but only at the 
expense of nullifying $\tilde\xi_0$, which is also unphysical).

The stationary in space-time covariance function $\gamma(t,s)$ of the random field $\xi(t,s)$
 can be easily derived 
from Eq.(\ref{xispe}) while utilizing the independence
of the  spectral processes $\tilde\xi_m(t)$, 
the spectral-space temporal covariance functions given in Eq.(\ref{Bmt}), and
returning to the Eulerian frame of reference:
\begin {equation}
\label{Btx}
\gamma(t,s) = \sum_{m=-n/2+1}^{n/2} b_m  \, \e^{ -{|\mathnormal{t}|}/\mathnormal{\tau_m}  } 
           \,\e^{\i \mathnormal{m {(s - Ut)}/R}}.
\end {equation}
Note that Eq.(\ref{Btx}) implies that the space-time correlations are {\em non-separable},
\ie they cannot be represented as a product of purely spatial and purely temporal correlations.
Moreover, according to Eq.(\ref{tau_spe}), 
smaller spatial scales (\ie larger wavenumbers $m$)
correspond to smaller time scales $\tau_m$.
This feature of space-time correlations (``proportionality of scales'')
is physically reasonable---as opposed to the simplistic and unrealistic 
separability of space-time correlations---and widespread in the real world, 
see  \citet[][]{PS,TsyGaySPG} and references therein. 

Finally, from Eq.(\ref{Btx}), the stationary (steady-state) variance of $\xi(t,s)$ is 
\begin {equation}
\label{var}
 \Var\xi \equiv (\sd(\xi))^2 = \sum_{m=-n/2+1}^{n/2} b_m =
          \frac {a^2 \sigma^2} {2} \sum_{m=-n/2+1}^{n/2}  \frac {1} {\rho + \frac {\nu} {R^2} m^2},
\end {equation}
where $\sd$ stands for standard deviation.

\subsection {Roles of model parameters}
\label{sec_roles}

Firstly, we note that $U$ does not impact the variance spectrum and the 
spectral time  scales of $\xi$
(see Eqs.(\ref{bm}) and (\ref{tau_spe})),
its role is just to rotate the solution with the constant angular velocity $U/R$.

Secondly, Eq.(\ref{bm}) implies that the {\em shape} of the spatial spectrum is 
\begin {equation}
\label{spe_shape}
 b_m \propto  \frac {1} {\rho + \frac {\nu} {R^2} m^2} 
               \propto   \frac {1} { 1 + (\frac {m} {m_0})^2 },
\end {equation}
where $m_0=R \sqrt{{\rho}/{\nu}}$ is the characteristic non-dimensional wavenumber, which
defines the {\em width} of the spectrum and thus  the field's length scale.
Therefore, the latter can be defined\footnote{
One can show that for a dense enough spatial grid,  thus defined length scale $L$
almost coincides with the macroscale $\Lambda_\xi$ defined below in Eq.(\ref{Lambda}).
}
  as the inverse dimensional wavenumber 
$m_0/R$:
\begin {equation}
\label{L}
L=\sqrt{\frac{\nu}{\rho}}.
\end {equation}
Thus, the ratio ${\nu}/{\rho}$ controls the length scale $L$.
In addition, ${\nu}/{\rho}$ impacts the temporal correlations.
Indeed, a higher $L$ implies a redistribution of the variance towards larger spatial scales (\ie lower wavenumbers $m$).
But as we noted, in the model Eq.(\ref{mdl}), 
larger spatial scales correspond to larger time scales $\tau_m$. 
As a result, a higher  ${\nu}/{\rho}$ leads to a larger  time  scale
as well as the length scale $L$.

Thirdly, using Eq.(\ref{L}), we can rewrite Eq.(\ref{tau_spe}) as 
\begin {equation}
\label{tau_spe2}
 \tau_m = \frac{1} {\rho} \cdot  \frac{1}{1 + (\frac {Lm} {R})^2}
\end {equation}
This equation implies that with $L$ being fixed, all spectral time  scales $\tau_m$
are inversely proportional to $\rho$, which, thus, determines the physical-space Lagrangian
time scale $T$ of the spatiotemporal random field $\xi$.
We define $T$ as the macroscale \citep[e.g.][Eq.(2.88)]{Yaglom} along the characteristic,
\begin {equation}
\label{T}
T=\frac{1}{2\Var\xi} \int_{-\infty}^\infty \gamma(t,Ut) \d t =
  \frac{ \sum b_m  \tau_m }{ \sum  b_m } =
  \frac{1}{\rho} \,
  \frac{ \sum [ 1 + (\frac {Lm} {R})^2 ]^{-2} } 
       { \sum [1 + (\frac {Lm} {R})^2]^{-1}  },
\end {equation}
where the second equality is due to Eq.(\ref{Btx}), 
the third equality is due to Eqs.(\ref{bm}) and (\ref{tau_spe2}), and 
the summations are over $m$ from $-\frac{n}{2}+1$ to $\frac{n}{2}$.

Technically,  Eqs.(\ref{L}), (\ref{T}), and (\ref{var}) 
allow us to compute the {\em internal} model parameters $\rho$, $\nu$, and $\sigma$ from the {\em externally}
specified parameters $L$, $T$, and $\sd(\xi)$.
Conceptually, we summarize the above conclusions as follows.
\begin{itemize}
\item
  $U$ does not affect the Lagrangian spatiotemporal covariances. It tilts the 
  Eulerian spatiotemporal correlations towards the direction $\d s =U \d t$ in space-time.

\item
 The ratio ${\nu}/\rho$ determines the spatial scale $L$ and impacts the time  scale $T$.

\item
With the ratio  ${\nu}/\rho$  being fixed, $\rho$ controls the time scale $T$.

\item
With $\rho$ and $\nu$ being fixed, $\sigma$ determines the resulting process variance $\Var\xi$.
\end{itemize}
These conclusions give us an idea how {\em local} properties
of the spatiotemporal field statistics are going to be impacted if the parameters 
$\boldsymbol\theta=(U,\rho,\nu,\sigma)$
become variable in space and time.

\section{Doubly stochastic advection-diffusion-decay model (DSADM)}
\label{sec_doub}

Here, we  allow the parameters $\boldsymbol\theta=(U,\rho,\nu,\sigma)$ of the model Eq.(\ref{mdl})    
to be spatiotemporal random fields by themselves.
The resulting model becomes, thus, a three-level hierarchical model \citep[][]{Wikle,Banerjee}.
At the first level is the random field in question $\xi(t,s)$ modeled conditionally on
the second-level fields $\boldsymbol\theta(t,s)=(U(t,s),\rho(t,s),\nu(t,s),\sigma(t,s))$,
which are 
controlled by the third-level hyperparameters $\boldsymbol\phi$.
So, to compute a realization of the pseudo-random field $\xi(t,s)$, we first
specify the hyperparameters $\boldsymbol\phi$.
Then, we  compute
realizations of the  second-level (secondary or parameter) fields $\boldsymbol\theta(t,s)$. 
Finally, we substitute the secondary fields for the respective parameters  in Eq.(\ref{mdl})
and solve the resulting equation for the primary field $\xi(t,s)$.

The idea behind this extension of the basic model  Eq.(\ref{mdl}) is the following.
If the secondary fields $\boldsymbol\theta(t,s)$
vary {\em smoothly} in space and time, then, locally, in a vicinity of some point in space-time $(t_0,s_0)$,
 the statistics of the field $\xi(t,s)$ will resemble 
that for the stationary model  Eq.(\ref{mdl}) with constant parameters equal to  $\boldsymbol\theta(t,s)$
{\em frozen} at the point $(t_0,s_0)$ \citep[see also][sec. 3.2]{Lindgren}.
As the statistics of the model Eq.(\ref{mdl}) with constant parameters do depend on the parameters 
$\boldsymbol\theta$ (see section \ref{sec_hom}), 
the resulting solution $\xi(t,s)$ to the model Eq.(\ref{mdl}) with variable parameters
 becomes non-stationary in space-time, 
with the degree of the non-stationarity controlled by the variability in the secondary fields 
$\boldsymbol\theta(t,s)$.

\subsection{First level of the hierarchy: the field in question $\xi$}

At the first level,  $\xi(t,s)$ satisfies 
the basic  Eq.(\ref{mdl}) with variable in space and time coefficients,
\begin {equation}
\label{dsm1}
\frac{\partial\xi} {\partial t} + U(t,s)  \, \frac{\partial \xi} {\partial s}  + \rho(t,s) \, \xi -
   \nu(t,s) \,  \frac{\partial^2\xi} {\partial s^2} =  
    \sigma(t,s) \,  \alpha(t,s). 
\end {equation}
%

\subsection{Second level of the hierarchy: parameter fields $\boldsymbol\theta$}
\label{sec_secondary}

Each secondary field $\theta(t,s)$ (that is, one of  the coefficients 
$U(t,s)$, $\rho(t,s)$, $\nu(t,s)$,  $\sigma(t,s)$ of the first-level Eq.(\ref{dsm1}))
is modeled as the {\em transformed  Gaussian field}: $\theta = g_\theta(\theta^*,\boldsymbol\phi)$.
Here $g_\theta$ is the (secondary-field specific) transformation function  and 
$\theta^*(t,s)$ is  the pre-transform Gaussian random field satisfying
its own stochastic advection-diffusion-decay model  Eq.(\ref{mdl}) with constant and 
non-random coefficients,
$U_\theta, \rho_\theta, \nu_\theta, \sigma_\theta$ (which are hyperparameters).

The pointwise transformation $g_\theta: \theta^*(t,s) \mapsto \theta(t,s)$ 
is specified to be linear for $U$ and nonlinear for the other three parameter fields.
The transformation function involves additional hyperparameters:
the ``unperturbed'' value of $\theta$ (a scalar) denoted  by the overbar, 
$\overline{\theta}$ (such that 
$\theta(t,s) = \overline{\theta}$ if $\Var\theta^*=0$), and
a few additional hyperparameters as described below.

Since the pre-transform fields $\theta^*(t,s)$ are governed by the models with {\em constant} coefficients,
$\theta^*(t,s)$ are stationary in space-time.
The transforms $g_\theta$ are defined to be independent of $(t,s)$, therefore  
the secondary fields $\theta(t,s)$ are stationary in space-time, too.

Next, the computation of the four  secondary  fields is described, 
including equations for the pre-transform random fields and 
the respective transformation functions.

\subsubsection{$U(t,s)$}

The pre-transform Gaussian field $U^*(t,s)$ satisfies
the basic  stochastic  model Eq.(\ref{mdl}):
\begin {equation}
\label{Umdl}
\frac{\partial U^*} {\partial t} +  U_U \frac{\partial U^*} {\partial s} + \rho_U U^* -
    \nu_U \frac{\partial^2 U^*} {\partial s^2} =  \sigma_U \, \alpha_U(t,s),
\end {equation}
where $U_U, \rho_U, \nu_U$, and $\sigma_U$ are 
constant and non-random hyperparameters
and $\alpha_U$ is the  white noise independent of $\alpha$.
The transformation $U^*(t,s) \mapsto U(t,s)$ is simply
\begin {equation}
\label{Ue}
U(t,s) = \overline U + U^*(t,s),
\end {equation}
where $\overline U$ is the unperturbed value of $U$ (a non-random scalar). 

From linearity of Eqs.(\ref{Umdl}) and (\ref{Ue}), $U(t,s)$ is a Gaussian random field.

\subsubsection{$\sigma(t,s)$}
\label{sec_sigma}

The pre-transform Gaussian field $\sigma^*(t,s)$ satisfies
\begin {equation}
\label{sigma_star}
\frac{\partial \sigma^*} {\partial t} + U_\sigma \frac{\partial \sigma^*} {\partial s} +\rho_\sigma \sigma^* -
    \nu_\sigma \frac{\partial^2 \sigma^*} {\partial s^2} =  \sigma_\sigma \, \alpha_\sigma(t,s).
\end {equation}
where
$U_\sigma, \rho_\sigma, \nu_\sigma$, and $\sigma_\sigma$ are the hyperparameters
and $\alpha_\sigma$ is the independent white noise field.

\begin{figure}
\begin{center}
   { 
   \scalebox{0.5}{ \includegraphics{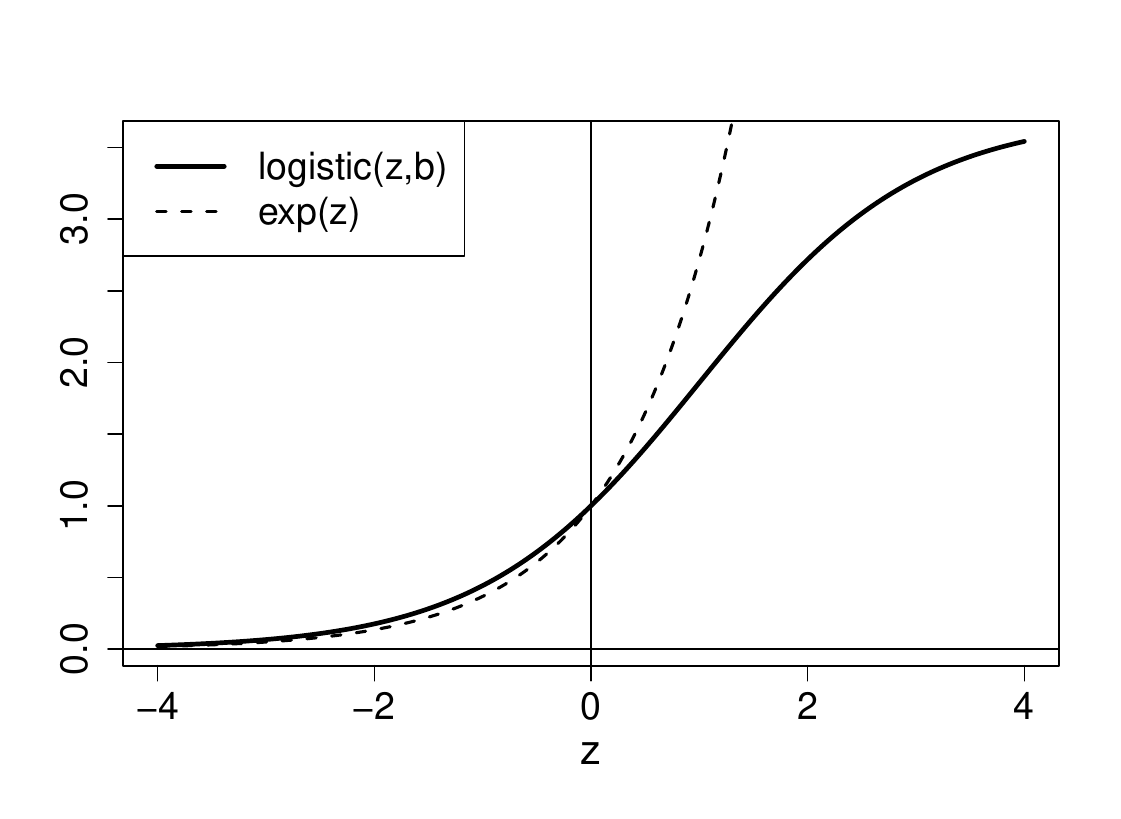}}
       }
\end{center}
  \caption{Logistic function  $g(z)$  with $b=1$ and exponential function}
\label{Fig_Logistic}
\end{figure}

To define the transformation $\sigma^*(t,s) \mapsto \sigma(t,s)$, we
require that the field $\sigma(t,s)$ should be  positive and 
have zero probability density at $\sigma=0$. To meet these requirements,
we postulate that
\begin {equation}
\label{sigma}
\sigma(t,s) = \overline\sigma \cdot g(\sigma^*(t,s)),
\end {equation}
where $\overline\sigma$ is the unperturbed value of $\sigma$ and
 $g(z)$ is the transformation function selected to be the  scaled and shifted  logistic function
(also known as the sigmoid function in machine learning):
\begin {equation}
\label{logist}
g(z) :=  \frac {1+\e^\mathnormal{b}} {1+\e^\mathnormal{b-z}},
\end {equation}
where $b$ is the constant. The function $g(z)$ has the following property:
it behaves like the ordinary exponential function everywhere except for $z \gg b$, 
where the exponential growth  is tempered (moderated).
Indeed, it exponentially decays as $z \to -\infty$.
Like $\exp(z)$, it is equal to 1 at $z=0$.
With $b >0$, $g(z)$ saturates as $z \to \infty$ at the level $1+ \e^\mathnormal{b}$;
this is the main difference of $g$
from the  exponential function and the reason why we replace $\exp(z)$ by $g(z)$:
to avoid too large values in  $\sigma(t,s)$, which can give rise to unrealistically large spikes in $\xi$.
We will refer to $b$ as the $g$-function's saturation hyperparameter.
For $b=1$, the function 
$g(z)$ in plotted in Fig.\ref{Fig_Logistic} alongside the exponential function.

Due to  nonlinearity of the transformation function $g$, the field $\sigma(t,s)$ is non-Gaussian.
Its pointwise distribution is known as logit-normal or logit-Gaussian.

\subsubsection{$\rho(t,s)$ and $\nu(t,s)$}

The remaining two secondary fields $\rho(t,s)$ and $\nu(t,s)$ (denoted here generically by $\psi$)
are modeled in a way similar to $\sigma(t,s)$, the only difference being the transformation function.
Specifically, the pre-transform Gaussian field $\psi^*$ satisfies
\begin {equation}
\label{psi_star}
\frac{\partial \psi^*} {\partial t} + U_\psi \frac{\partial \psi^*} {\partial s} +\rho_\psi \psi^* -
    \nu_\psi \frac{\partial^2 \psi^*} {\partial s^2} =  \sigma_\psi \, \alpha_\psi(t,s),
\end {equation}
where, again, $U_\psi, \rho_\psi, \nu_\psi$, and $\sigma_\psi$, are the hyperparameters
and $\alpha_\psi$ is the independent white noise field.

The transformation function is defined here to be
\begin {equation}
\label{psi}
\psi(t,s) = \overline\psi \cdot \left[ (1 + \varepsilon_\psi) \cdot g(\psi^*(t,s))-  \varepsilon_\psi \right],
\end {equation}
where  $\overline\psi$ is the unperturbed value of $\psi$, 
the function $g$ is the same as in section \ref{sec_sigma} 
and controlled by the same saturation hyperparameter $b$, and
$\varepsilon_\psi$ is the small non-negative constant.
The hyperparameter $\varepsilon_\psi$ is introduced to allow for
small negative values of $\psi$ (that is, of  $\rho$ and $\nu$).
This will take place if the pre-transform Gaussian field $\psi^*$ happens to take a large negative value.
Allowing for  negative values of  the decay coefficient $\rho$ and the diffusion coefficient $\nu$ is motivated by the desire to 
introduce an intermittent instability into the model.

Like $\sigma(t,s)$, the fields $\rho(t,s)$ and $\nu(t,s)$ are logit-Gaussian.

\subsection {Third level of the hierarchy: hyperparameters and external parameters}
\label{sec_intext}

In contrast to the popular  Lorenz'96 model \citep{Lorenz1998}, which has only one parameter $F$,  DSADM has many parameters
as discussed above (in total, there are 23 hyperparameters).
It is  convenient to specify the   hyperparameters using a set of
more sensible {\em external} parameters as described next.

The advection velocities $\overline U$ and $U_\theta$ (for $\theta=U,\rho,\nu,\sigma$)
are specified directly as they have a clear physical meaning.
 
The unperturbed hyperparameters $\overline\theta=\overline\rho, \overline\nu, \overline\sigma$
 are specified from the unperturbed external parameters $\overline L$, $\overline T$, and $\overline{\sd(\xi)}$
  using Eqs.(\ref{T}), (\ref{L}), and (\ref{var})\footnote{Note that for any field $\theta$, $\overline{\theta}$
is its pointwise {\em median} (for $\theta=U$ it is also the mean), hence the overbar notation.}.
The latter determine the desired typical  spatial scale, time scale, and variance of the field $\xi(t,s)$.
Note that  in the non-stationary regime, $\overline L$, $\overline T$, and $\overline{\sd(\xi)}$ 
are not exactly equal to the mean values of the length scale, time scale, and variance of $\xi(t,s)$,
respectively, because closed-form expressions for the latter are not available.

Likewise,  for any pre-transform Gaussian field $\theta^*= U^*, \rho^*, \nu^*, \sigma^*$,
the respective hyperparameters $\rho_\theta$ and $\nu_\theta$ are calculated from  the 
external parameters $L_\theta$ and  $T_\theta$ using  Eqs.(\ref{T}) and (\ref{L}).
$L_{\theta}$ and $T_{\theta}$ are 
 the  user-specified {\em spatial and temporal scales of non-stationarity}.

The  hyperparameters  $\sigma_\theta$ are specified as follows.  $\sigma_U$ is calculated from
the external parameter $\sd (U^*)$ using Eq.(\ref{var}).
For each of the other three secondary fields, $\theta=\rho,\nu,\sigma$, we select $\varkappa_\theta = \exp(\sd(\theta^*))$ 
to be the respective external parameter. 
This  choice is motivated by the fact that these fields are nonlinearly transformed Gaussian fields. 
As $g(z)$ (defined in Eq.(\ref{logist}) and shown in Fig.\ref{Fig_Logistic}) 
is a ``tempered'' exponential function, it is worth measuring the standard deviation of,
say, the $\sigma^*$ field 
on the log scale: $\sd(\sigma^*)=\log\varkappa_\sigma$, so that the typical deviation of 
the transformed field $\sigma$  from its unperturbed value $\overline\sigma$ is $\varkappa_\sigma$  {\em times}.
The external parameters  $\sd (U^*)$, $\varkappa_\sigma$, $\varkappa_\rho$, and $\varkappa_\nu$ 
determine the {\em strength} of non-stationarity.

Finally, for the fields $\psi=\rho,\nu$,  we parameterize $\varepsilon_\psi$ 
(which, we recall, controls the occurrence of {\em negative} values of $\psi$) using  the
probability $\pi_\psi$ that $\psi(t,s)$ is  negative:
\begin {equation}
\label{Pneg}
P(\psi(t,s) < 0)= \pi_\psi.
\end {equation}
Substituting $\psi$ from Eq.(\ref{psi}) into Eq.(\ref{Pneg}) and utilizing  monotonicity of 
the transformation function $g$ and  Gaussianity of the field $\psi^*$, we easily
come up with a relation between $\pi_\psi$ and $\varepsilon_\psi$ (the elementary formulas are
omitted).
The external parameters $\pi_\rho$ and $\pi_\nu$ thus determine how often and how strong {\em local instabilities} can be.
(Actually, $\pi_\psi$ slightly impacts also  $\sup \psi$, but this is a minor effect and it can be ignored in this application.)

To facilitate the use of DSADM, we 
introduce a reduced set of external parameters using the following constraints.

(i) The unperturbed advection velocity  $\overline U$ and all advection velocities $U_{\theta}$
 are equal to each other.

(ii) The length-scale hyperparameters $L_{\theta}$ for all  pre-transform fields $\theta^*$
are selected to be equal to the common value $L^*$, the {\em length scale of  non-stationarity}.

(iii) The time-scale  hyperparameters $\overline T$ and $T_{\theta}$ are specified
to be equal to $\overline L/V_{\rm char}$ and $L_{\theta}/V_{\rm char} = L^*/V_{\rm char}$, respectively, 
where $V_{\rm char}$  is the {\em characteristic velocity}.

(iv) The $g$-function's saturation hyperparameter $b$ is set to 1.

Having selected a meaningful value of $V_{\rm char}$, the user can control the 
typical time and length scales of $\xi$, the strength of  non-stationarity, and the time 
and length scales  of  non-stationarity --- using the reduced set of the 10 external parameters 
listed  in Table \ref{Tab_params}.

\begin{table}[ht]
\caption{Short list of external parameters of DSADM}

\begin{center}
\begin{tabular}{|l|l|l|}

\hline
External  & Dependent &  Which feature of the model is controlled?\\
 parameter      & hyperparameters & \\
\hline
\hline
$\overline{\sd(\xi)}$     &       $\overline\sigma$              &  Mean $\sd(\xi)$ (roughly)   \\
\hline
$\overline U$ &    $\overline U, U_{U}, U_\rho, U_\nu, U_\sigma$       &  All advection velocities   \\
\hline
$\overline L$            &   $\overline\rho,\overline\nu$   &  Mean length scale of $\xi$ (roughly)   \\
\hline
$L^*$                &  $\rho_U, \rho_\rho, \rho_\nu, \rho_\sigma$     &  Length scale and time scale\\
                        &  $\nu_U, \nu_\rho, \nu_\nu, \nu_\sigma$     & of non-stationarity   \\
\hline
$\sd (U^*)$           &    $\sigma_U$  &    Strength of non-stationarity \\
 \hline          
$\varkappa_\rho, \varkappa_\nu, \varkappa_\sigma$ & $\sigma_\rho, \sigma_\nu, \sigma_\sigma$& Strength of non-stationarity\\             
 \hline              
 $\pi_\rho$, $\pi_\nu$     & $\varepsilon_\rho$, $\varepsilon_\nu$ & Strength of non-stationarity \\   
               
      &  & (portions of time $\rho$ and $\nu$ are negative)\\   
  \hline          
\hline
\end{tabular}
\end{center}
\label{Tab_params}
\end{table}
%

\section {Properties and capabilities of DSADM}
\label{sec_propert}

\subsection {Non-stationarity}
\label{sec_truecovs}

Here, we show that, given the secondary fields, the solution to DSADM is indeed 
a non-stationary in space and time  random field. 
Its spatiotemporal  covariances are themselves random (as they depend on the 
random secondary fields) and appear to be stationary processes in space-time.

Let us rewrite the basic Eq.(\ref{dsm1}) as the linear stochastic state-space  model
\begin {equation}
\label{dxidt}
\frac{d\boldsymbol\xi(t)} {\d t} = \boldsymbol\Phi(\underline{\boldsymbol{\theta}}(t)) \, \boldsymbol\xi(t) + 
     \boldsymbol\Sigma(\underline{\boldsymbol{\theta}}(t)) \, \boldsymbol\alpha(t),
\end {equation}
where  $\boldsymbol\xi$ and $\underline{\boldsymbol{\theta}}$  stand for the vectors of the 
spatially gridded fields $\xi(t,s)$ and $\boldsymbol{\theta}(t,s)$, respectively, $\boldsymbol\alpha(t)$
is the space-discrete and time-continuous white noise,
$\boldsymbol\Phi(\underline{\boldsymbol{\theta}})$
is the spatial operator dependent on the spatial fields $\underline{\boldsymbol{\theta}}$, and
$\boldsymbol\Sigma(\underline{\boldsymbol{\theta}})$ is the diagonal matrix whose application 
to $\boldsymbol\alpha$ represents the forcing term $\sigma \alpha$ in Eq.(\ref{dsm1}).	

Note that DSADM is intended to be used as a model-of-truth in which the  secondary fields,
once computed in an experiment, are held fixed.
In this default setting, Eq.(\ref{dxidt})  implies that DSADM is a linear non-autonomous stochastic model
(because $\boldsymbol\Phi$ and $\boldsymbol\Sigma$ are explicit functions of time).

Discretizing Eq.(\ref{dxidt}) in time 
(using an implicit time-differencing scheme)
yields the equation for the time-discrete signal $\boldsymbol\xi$:
\begin {equation}
\label{dxidt_dscr}
\boldsymbol\xi_k  = {\bf F}(\underline{\boldsymbol{\theta}}_k) (\boldsymbol\xi_{k-1} + 
                              \boldsymbol\Sigma(\underline{\boldsymbol{\theta}}_k) \boldsymbol\alpha_k),
\end {equation}
where $\boldsymbol\xi_k$ is the field on the spatial grid (an $n$-vector),
$k=1,2,\dots$ labels the time instant,
${\bf F}$ is the model operator, and components of the vector
$\boldsymbol\alpha_k$  are independent Gaussian variables with mean zero and variance $1/(\Delta s \Delta t)$.
Here $\Delta s=2\pi R/n$ is the spatial grid spacing and $\Delta t$ the time step.
The initial condition $\boldsymbol\xi_0$ has mean zero and the known 
covariance matrix $\boldsymbol\Gamma_0$.
Equation (\ref{dxidt_dscr}) allows us to find the conditional covariance matrix of the spatial field $\boldsymbol\xi_k$
given the secondary fields $\underline{\boldsymbol{\theta}}_{1:k}$ 
(the notation $1:k$ designates all time instants from 1 to $k$),
that is, 
$\boldsymbol\Gamma_k = \Ex (\boldsymbol\xi_k \boldsymbol\xi_k^\top | \underline{\boldsymbol{\theta}}_{1:k})$,
using the recursion
\begin {equation}
\label{Bk}
\boldsymbol\Gamma_k  = {\bf F}(\underline{\boldsymbol{\theta}}_k) \, \boldsymbol\Gamma_{k-1} \,  
                                    {\bf F}(\underline{\boldsymbol{\theta}}_k)^\top +
                                    {\bf Q}(\underline{\boldsymbol{\theta}}_k),
\end {equation}
where ${\bf Q} = {\bf F} \boldsymbol\Sigma^2 {\bf F}^\top /(\Delta s \Delta t)$ 
is the covariance matrix of the  forcing.
The lagged  covariances $\boldsymbol\Gamma_{kj} = 
\Ex (\boldsymbol\xi_k \boldsymbol\xi_j^\top | \underline{\boldsymbol{\theta}}_{1:k})$
(where $j<k$) can be found by right-multiplying Eq.(\ref{dxidt_dscr}) by $\boldsymbol\xi_j^\top$
and taking expectation (note that for $j<k$, $\boldsymbol\xi_j$ is independent of $\boldsymbol\alpha_k$):
\begin {equation}
\label{Glag}
\boldsymbol\Gamma_{kj}  = {\bf F}(\underline{\boldsymbol{\theta}}_k) \, \boldsymbol\Gamma_{k-1,j}.
\end {equation}
Now recall that the secondary fields $\underline{\boldsymbol\theta}$ are random fields.
Then, Eqs.(\ref{Bk}) and (\ref{Glag}) imply that  the field (signal) covariances  $\boldsymbol\Gamma_k$ and 
$\boldsymbol\Gamma_{kj}$ are (matrix variate) random processes.
It can be shown that if  $\rho(t,s)>0$ and $\nu(t,s)>0$ 
(for small enough $\pi_\rho$ and $\pi_\nu$ this is true most of the time), 
then  ${\bf F}$ (as a matrix operator) 
is a contraction. Therefore,  the dependence of $\boldsymbol\Gamma_k$ 
and  $\boldsymbol\Gamma_{kj}$ on $\boldsymbol\Gamma_0$
fades out as $k\to \infty$ and only the secondary fields 
$\underline{\boldsymbol\theta}$ determine the covariances of $\xi(t,s)$.
Moreover, since the fields $\underline{\boldsymbol\theta}$ are stationary in space-time
and  ${\bf F}$ and ${\bf Q}$  are space and time invariant
(as  operators acting on $\underline{\boldsymbol\theta}$), 
the field covariances $\boldsymbol\Gamma_k$ and 
$\boldsymbol\Gamma_{kj}$ are stationary random processes in space-time.

Below, we will be particularly interested in the following two aspects of 
the spatial covariances $\boldsymbol\Gamma_k$ or, in  continuous notation and omitting for brevity 
the dependence on the secondary fields, $\gamma(t,t;s,s')$:
(i) the time and space specific {\em field variance} 
$\Var\xi(t,s)=\gamma(t,t;s,s)$
and (ii) a time and space specific
{\em spatial scale} defined to be, say, the local macroscale:
\begin {equation}
\label{Lambda}
\Lambda_\xi(t,s) = \frac{1}{2\, \Var\xi(t,s)} \int_{\S^1(R)} \gamma(t,t; s,s') \d s'.
\end {equation}
Then, from stationarity of $\gamma(t,t;s,s')$, it follows that both $\Var\xi(t,s)$ and $\Lambda_\xi(t,s)$
are stationary in space-time random fields.

\subsection {Estimation of true covariances}
\label{sec_esttruecovs}

Instead of the recursive computation of the field covariances using Eq.(\ref{Bk}),
one can {\em estimate} them as accurately as needed by
running DSADM Eq.(\ref{dsm1}) $P$  times with independent realizations of the forcing  $\alpha(t,s)$
and with the fields $\boldsymbol{\theta}(t,s)$ held fixed.
The spatial covariances $\gamma(t,t;s_1,s_2)$
can then be estimated from the resulting sample $\{ \xi^{(p)}(t,s) \}_{p=1}^P$ as 
the usual sample covariances.

More importantly, this approach can be used to estimate  
time and space-specific 
{\em true filtering distributions} of any filter in question.
Note that these distributions and their parameters (\eg true background-error covariances)
are hard to obtain even knowing the  model of truth.
Indeed, on the one hand, a sub-optimal filter,
because of its approximate nature,  can only yield estimates which
are inexact and, likely, biased (hence the need for covariance inflation in EnKF).
On the other hand, an exact (\eg Kalman) filter 
cannot be used here either ---
because the error statistics of the approximate filter
and the exact filter differ.

To estimate, say,  true background-error covariances of a filter, 
one may use the above fields $\{ \xi^{(p)}(t,s) \}_{p=1}^P$ as $P$  ``truths'',
generate $P$ sets of ``synthetic'' observations 
(with independent errors)
in space and time, and perform $P$ assimilation runs.
Then, at each time, one may compute $P$ background-error vectors 
(by subtracting the truth $\xi^{(p)}(t,s)$ from the respective background field)
and then, finally, compute their (time specific) sample covariance matrix.
We experimented with DSADM on a 60-point spatial grid and found that the sample size 
$P=5000\mbox{--}10000$ was enough to accurately
estimate true time-specific error covariances  (not shown).
This approach is similar to that by \citet{bishop}; the difference is 
that in our approach the truth is random whilst they assume that the truth is deterministic.

\subsection {Instability}
\label{sec_instab}

The nonlinear deterministic models mentioned in the Introduction are chaotic, that is, having unstable modes (positive Lyapunov exponents),
whereas DSADM is stochastic but experiencing intermittent instabilities due to the possibility for $\rho$ and $\nu$
to attain negative values.
In the deterministic models instabilities are curbed
by nonlinearity, whereas in DSADM, these are limited by the time 
the random fields $\rho$ and $\nu$ remain negative.

\subsection {Gaussianity}
\label{sec_Gau}

Given $\boldsymbol\theta$,  DSADM is linear, hence $\xi(t,s)$ is conditionally Gaussian.
Unconditionally,  $\xi(t,s)$ is  a continuous mixture of 
zero-mean Gaussian distributions and so  has a
non-Gaussian distribution with heavy tails\footnote{
Indeed, one can prove using the Jensen inequality that the {\em kurtosis} of
a non-degenerate mixture of this kind (a scale mixture) is always greater than 3 (the Gaussian kurtosis).
A positive excess kurtosis means, normally, more probability mass in the tails of the distribution
than in the tails of the Gaussian distribution with the same mean and variance.
}. 

\subsection {Unbeatable benchmark filter}
\label{sec_bench}

Linearity of DSADM makes it possible to use the exact Kalman filter and thus to know how far
from the optimal performance  the filter in question is, which is always useful but often not possible with nonlinear 
deterministic models of truth.

In principle, non-stationarity combined with linearity can be achieved by
taking a nonlinear model (like those mentioned in the Introduction) and using its tangent linear model.
We have tried that with the Lorenz'96 model and found that even a tiny imposed 
model-error perturbation leads to an explosive growth of the forecast perturbation.
Perhaps, some diffusion is needed to stabilize the model.
This would complicate the model, 
but more significantly,
the  tangent linear model does not allow to explicitly specify the various characteristics of non-stationarity
the DSADM does.

\section {Behavior of the model}
\label{sec_behav}

In this section, we numerically examine solutions to DSADM and some aspects of their
spatiotemporal distributions.

\subsection {Model setup}
\label{sec_mdl_setup}

The model's differential equations were solved numerically using an 
implicit upwind finite-difference scheme.
The computational grid had $n=60$ points 
on the circle of radius $R=6370$ km.
The model integration time step was $\Delta t=6$h.
The external parameters of DSADM were selected to resemble the spatiotemporal structure
of a mid-tropospheric meteorological field like temperature or geopotential, with one caveat:
the time and length scales were chosen to be about twice as  large as the respective meteorological scales
(for the fields to be reasonably resolved on the  60-point spatial grid).

We specified
$\overline U=$ 10 ms$^{-1}$, 
$\overline L = 5 \Delta s = 3300$ km,
$L^*= 2 L$  (it is meaningful to assume that the {\em structural} change 
in the field occurs at a larger space and time scale than the change in the random field itself),
$\overline{\sd(\xi)}=5$ (selected arbitrarily and does not impact any conclusions),
$\sd (U^*)=10$ ms$^{-1}$,
$b=1$,
$\varkappa_\rho=\varkappa_\nu=\varkappa_\sigma=3$,
$\pi_\rho=0.02$,
$\pi_\nu=0.01$,   
$V_{\rm char}=3$ ms$^{-1}$  (tuned to give rise to the desired mean time scale).

\subsection {$\xi(t,s)$ plots}

\begin{figure}
\centering
  
   
   \sidesubfloat[][]{ \includegraphics[scale=0.4]{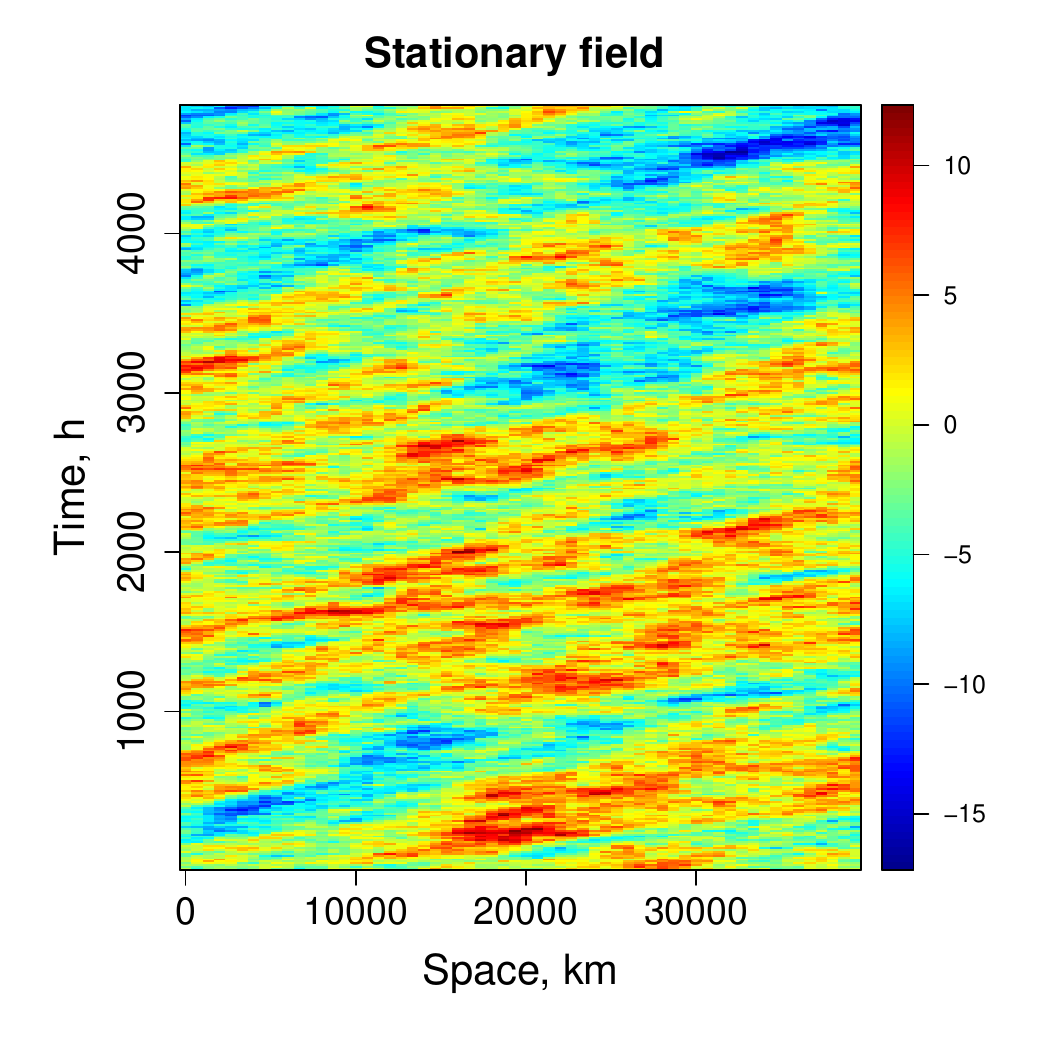}}
   \sidesubfloat[][]{ \includegraphics[scale=0.4]{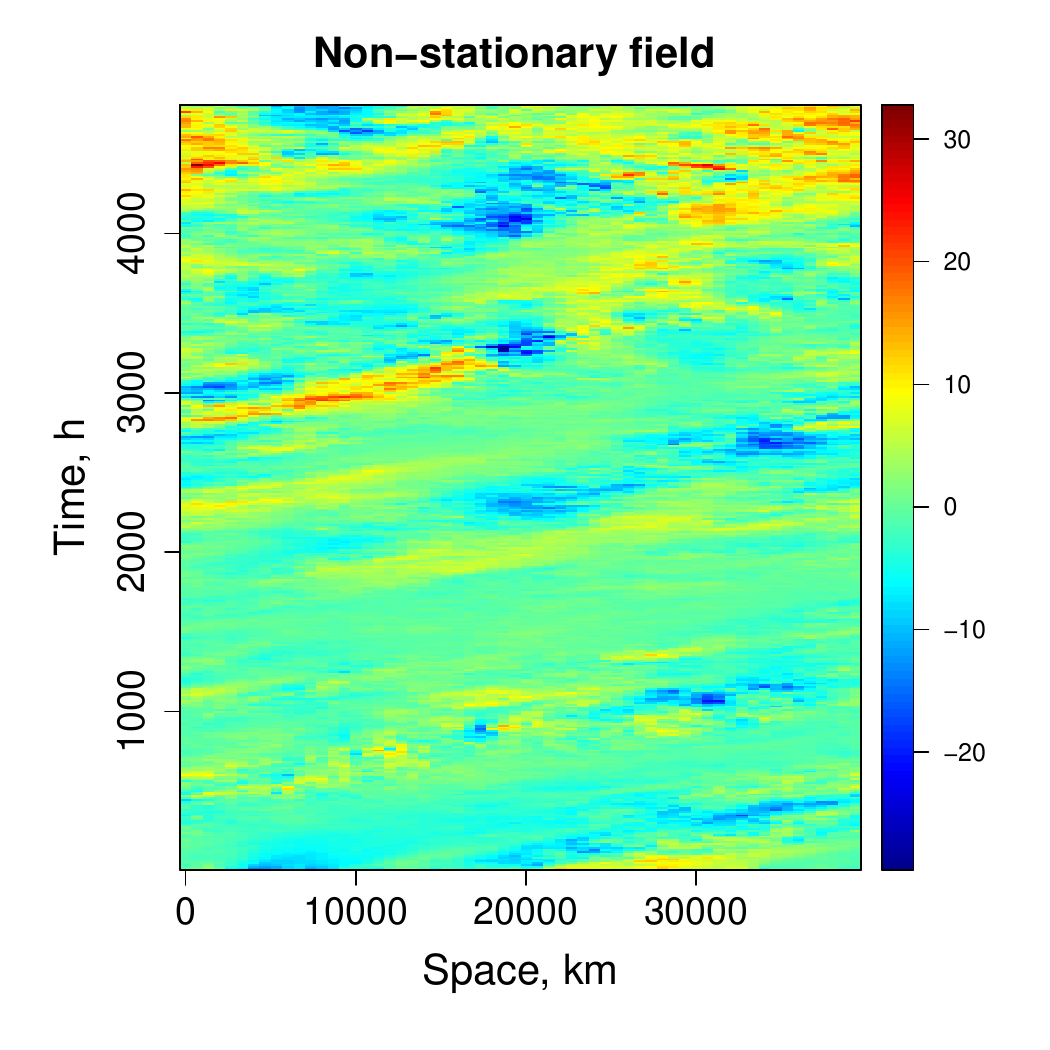}}
   
    \caption{Realizations of $\xi(t,s)$ generated by: 
         {(a)} the stationary  model Eq.(\ref{mdl}),
         {(b)} the non-stationary DSADM  Eq.(\ref{dsm1}). 
              }
\label{Fig_xi}
\end{figure}

Figure  \ref{Fig_xi}  compares typical spatiotemporal segments of solutions to
 the stationary stochastic model, Eq.(\ref{mdl}), (panel a)  and 
DSADM, Eq.(\ref{dsm1}),  (panel b).
One can see that, indeed, the non-stationary field 
had different structures at different areas across the domain,
whilst the stationary field had the same structure everywhere.
In particular, one can spot areas where the non-stationary field
experienced more small-scale (large-scale) fluctuation than in the rest of the plot. 
These spots correspond to small (large) values of the local length scale  $\Lambda_\xi$, see below 
Fig.\ref{Fig_VLambda}(b).

\subsection {Non-stationarity}
\label{sec_behav_Var_L}

\begin{figure}
\centering


   \sidesubfloat[][]{ \includegraphics[scale=0.4]{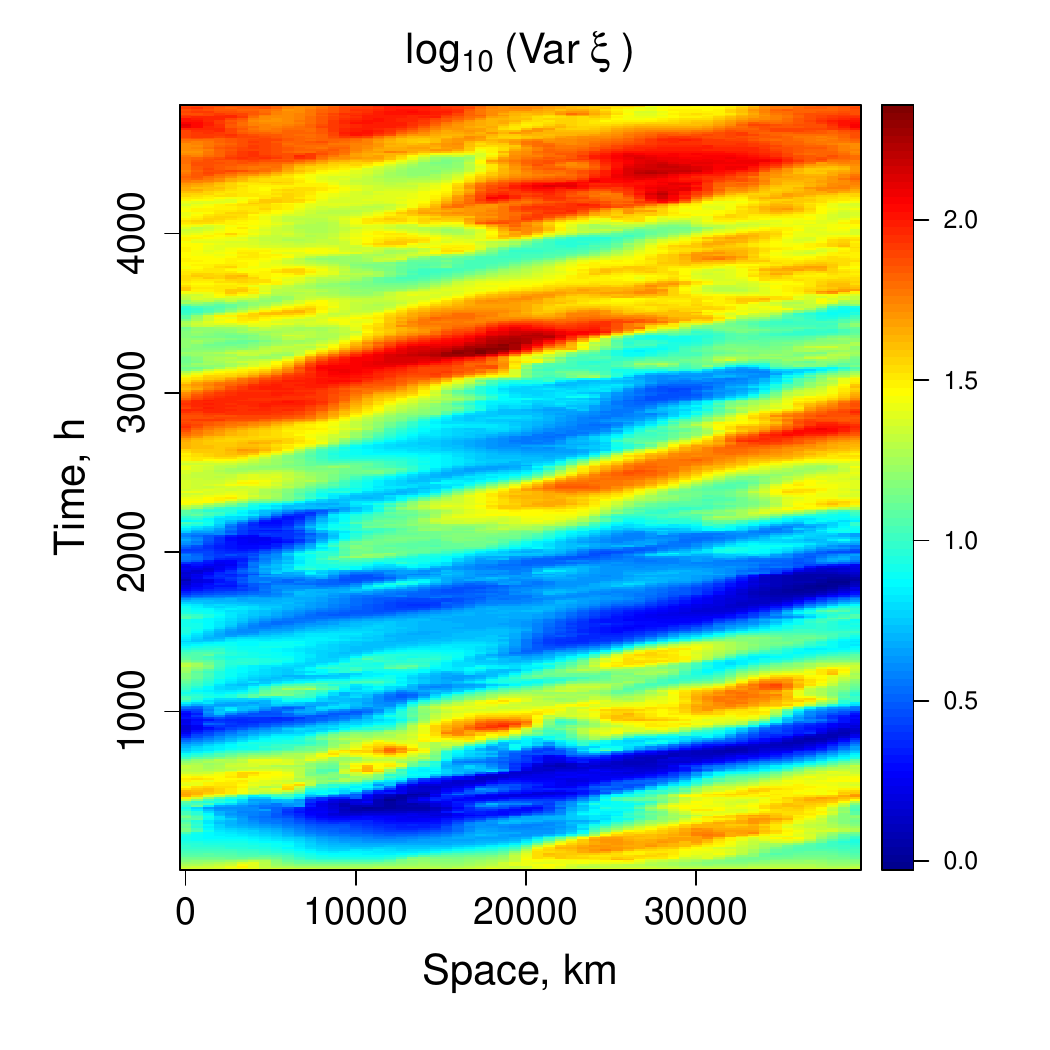}}
   \sidesubfloat[][]{ \includegraphics[scale=0.4]{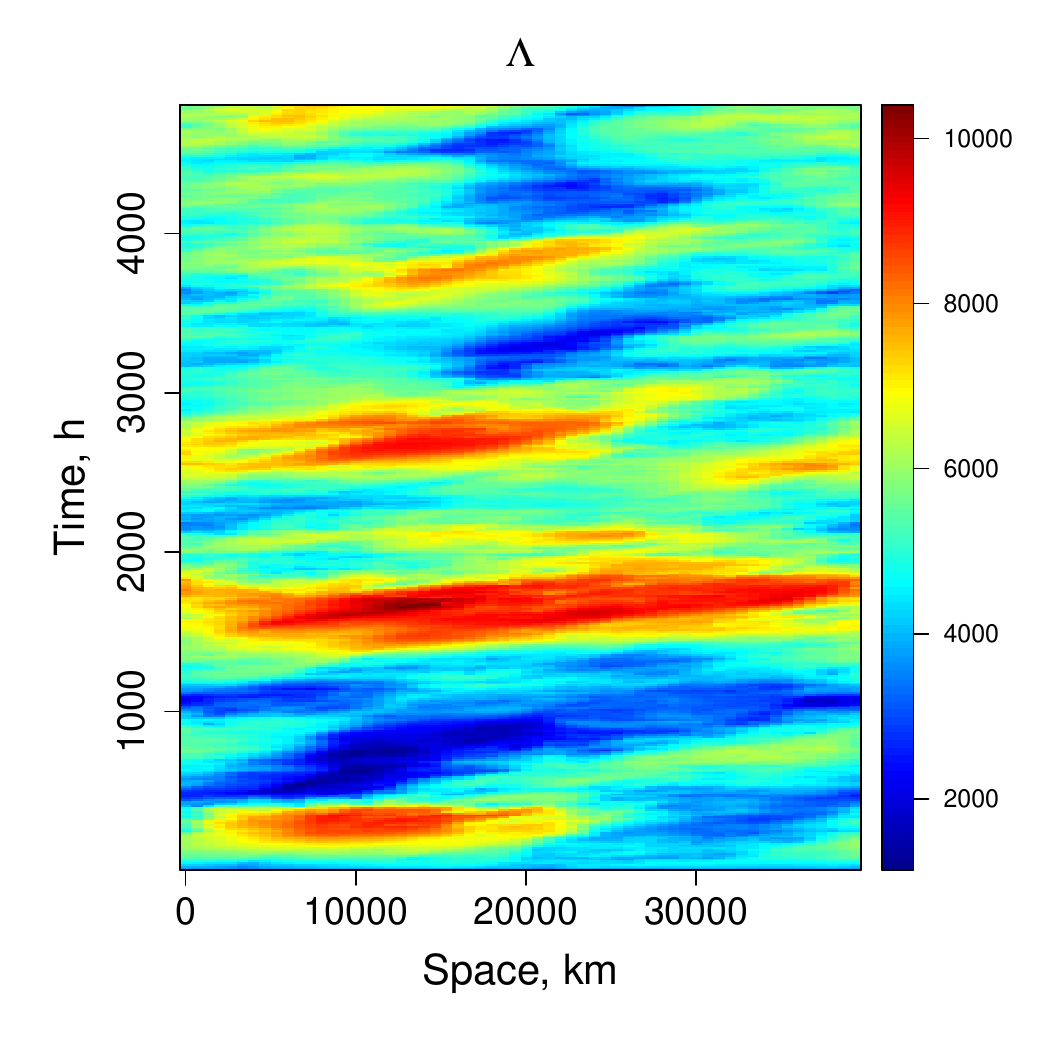}}

  \caption{Space-time plots of estimated
         {(a)}   field log-variance  $\log_{10} (\Var\xi(t,s))$,
         {(b)}   spatial macroscale $\Lambda_\xi(t,s)$.  
              }
\label{Fig_VLambda}
\end{figure}

Figure \ref{Fig_VLambda} shows two characteristics of  the non-stationarity pattern:
the local log-variance $\log_{10} (\Var\xi(t,s))$ 
and the local length scale $\Lambda_\xi(t,s)$ (defined in Eq.(\ref{Lambda})).
Both were  calculated from the spatial covariance matrix $\boldsymbol\Gamma_k$
computed following Eq.(\ref{Bk}). 
We recall that the variability in $\Var\xi(t,s)$ and $\Lambda_\xi(t,s)$ was induced 
by the simulated secondary  fields $\boldsymbol\theta(t,s)$,
on which ${\bf F}_k$ and ${\bf Q}_k$ depend.
The same realizations of the secondary fields were used to plot both Fig.\ref{Fig_VLambda} 
and Fig.\ref{Fig_xi}(b). 
In Fig.\ref{Fig_VLambda}, one can see the substantial degree of non-stationarity (note  that in the stationary case 
both $\Var\xi$ and $\Lambda_\xi$ are constant). 
Specifically, in Fig.\ref{Fig_VLambda}(a),
the ratio of the maximum to the minimum field variance $\Var \xi(t,s)$ is seen to be
greater than  two orders of magnitude.
The same ratio for the length scale $\Lambda_\xi(t,s)$ (see Fig.\ref{Fig_VLambda}(b)) was about 5, which also
indicates a significant degree of variation.

\begin{figure}
\begin{center}
   { 
      \scalebox{0.55}{ \includegraphics{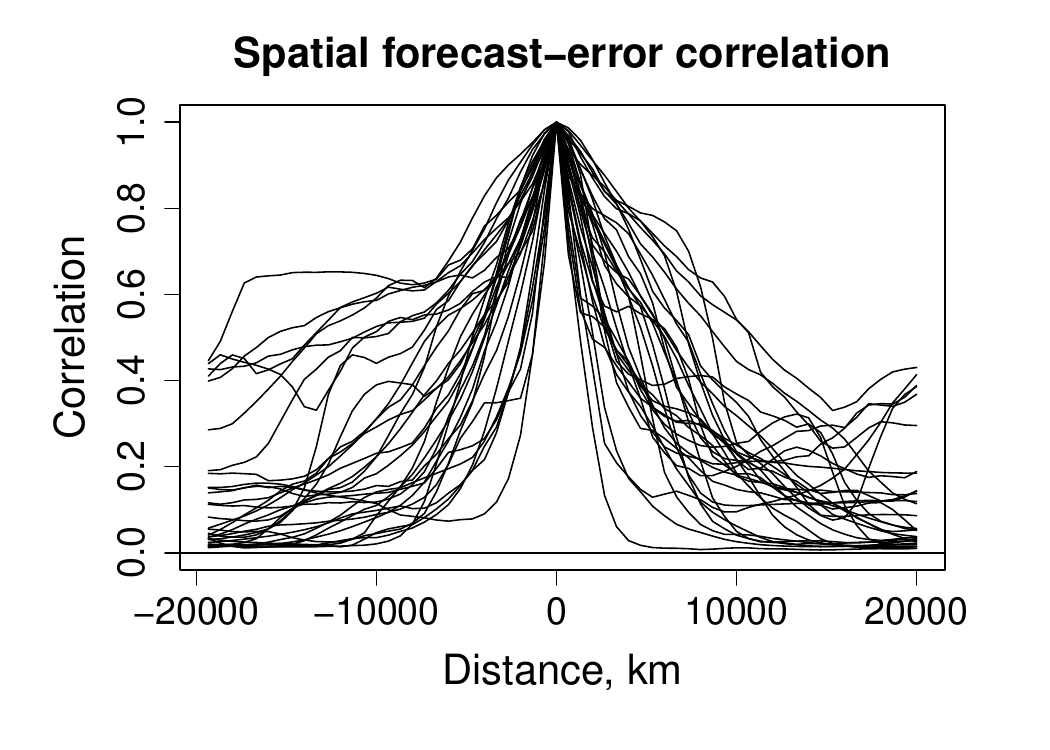}}
    }
\end{center}
  \caption{Spatial correlations of the field $\xi$ with respect to 30 randomly selected points in space-time.
              }
\label{Fig_crl}
\end{figure}

Figure \ref{Fig_crl} displays how diverse the true spatial {\em correlations} were.
Note that Fig.\ref{Fig_VLambda}(a) illustrates the non-stationarity of the field's magnitude,
whereas Figs.\ref{Fig_VLambda}(b) and \ref{Fig_crl}
highlight the non-stationarity of the field's spatial structure.

Thus,  DSADM is capable of generating significantly non-stationary random fields.

\subsection {Gaussianity}
\label{sec_behav_Gau}

As noted above, conditionally on the secondary fields,
$\xi(t,s)$ is Gaussian by construction.
This is the way DSADM is used in this study.
In principle,  it can be used for other purposes 
without conditioning on the secondary fields,
so that in each model run both the forcing and the model coefficients are random
(and the model becomes autonomous).
In this setting, the generated field  $\xi$ is no longer Gaussian.
 Numerical experiments confirmed that the unconditional
distribution of $\xi(t,s)$ was indeed non-Gaussian with  heavy tails 
(as we anticipated in section \ref{sec_Gau}).
The non-Gaussianity was stronger for larger 
magnitudes of the secondary fields (not shown).

\subsection {Forward and inverse modeling}
\label{sec_discu_dsadm}

Building DSADM
can be called  {\em forward} hierarchical modeling.
That is, we have formulated a (hopefully) reasonable hierarchical model and an algorithm
to compute realizations of the first-level random field $\xi(t,s)$ given the third-level
hyperparameters $\boldsymbol\phi$.
A harder problem is  {\em inverse} modeling, that is, 
the inference about the model parameters---in our case
the parameter fields $\boldsymbol\theta(t,s)$---from a number of realizations 
(an ensemble) of the field $\xi$.
This is the classical hierarchical Bayesian problem,
which is beyond the scope of this study but can be relevant 
in a broader context
of non-stationary spatial and spatiotemporal field modeling.

\section {Hybrid HBEF (HHBEF) filter}
\label{sec_hhbef}

In the rest of the paper, we use  DSADM to study the impact of  non-stationarity
on the performance of hybrid ensemble filters.
We formulate a new filter that extends HBEF (in which  ensemble 
covariances are blended with recent-past time-smoothed covariances, see
\citet{TsyRa}) by 
introducing two additional kinds of hybridization mentioned in the Introduction:
blending with climatological covariances (as in EnVar) and 
blending with spatially neighboring covariances (spatial smoothing). 
We start with a review of HBEF and then introduce HHBEF.

\subsection {HBEF}
\label{sec_hbef}

\citet{TsyRa} developed, building on \cite{Myrseth,Bocquet,Bocquet2015},  their
Hierarchical Bayes ensemble Kalman filter (HBEF), which cyclically
updates prior covariances using 
ensemble members as generalized observations and the estimated previous-cycle
covariances as a background.
The (Bayesian) update relies on the inverse-Wishart matrix-variate 
hyperprior distribution for the unknown, and thus assumed random, covariance matrices.

HBEF updates the model-error and the predictability-error covariance matrices,
but here we consider a simplified design in which the background-error covariance matrix ${\bf B}_k$
is cyclically updated.
In the simplest version of  HBEF, 
the mean-square-optimal posterior (analysis) estimate  ${\bf B}^{\rm a}_k$ 
of the unknown true matrix ${\bf B}_k$
is the linear combination of the {\em background} ${\bf B}^{\rm f}_k$ 
and the (localized and inflated) sample  covariance matrix, ${\bf B}^{\rm e}_k$:
\begin {equation}
\label{Ba}
{\bf B}^{\rm a}_k =  \frac {\vartheta  {\bf B}^{\rm f}_k + N {\bf B}^{\rm e}_k} {\vartheta +N},
\end {equation}
where 
$N$ is the ensemble size and $\vartheta >0$ the so-called sharpness parameter 
of the inverse Wishart distribution (defined in Appendix A in \citet{TsyRa}).
The background ${\bf B}^{\rm f}_{k}$
is provided by the  persistence forecast:
\begin {equation}
\label{Bf}
{\bf B}^{\rm f}_{k} =  {\bf B}^{\rm a}_{k-1}.
\end {equation}

The matrix ${\bf B}^{\rm a}_k$ computed according to Eq.(\ref{Ba}) 
is used in the analysis of the state $\boldsymbol\xi_k$
as the background-error covariance matrix.
The posterior (analysis) ensemble is generated in HBEF (and in all other filters in this study)
in the same way as in the stochastic EnKF \citep[e.g.,][]{Houtekamer2016review}.

From  Eqs.(\ref{Ba}) and (\ref{Bf}), it follows that ${\bf B}^{\rm a}_k$
satisfies the  first-order autoregressive equation
\begin {equation}
\label{tBsmo}
{\bf B}^{\rm a}_k = \mu {\bf B}^{\rm a}_{k-1} + (1-\mu) {\bf B}^{\rm e}_k,
\end {equation}
where $\mu = \vartheta/ (\vartheta +N)<1$. Equation (\ref{tBsmo}) 
implies  a kind of time smoothing of ensemble covariances.
In particular, ${\bf B}^{\rm a}_{k-1}$ is the time-smoothed
recent-past ensemble covariance matrix  denoted in what follows by  ${\bf B}^{\rm r}_k$.
With this notation, Eq.(\ref{tBsmo}) shows that ${\bf B}^{\rm a}_k$ is a 
(convex) linear combination of ${\bf B}^{\rm r}_k$ and ${\bf B}^{\rm e}_k$.
The role of ${\bf B}^{\rm r}_k$ in HBEF is two-fold. First, if  spatial non-stationarity  
has some memory (which is highly likely in realistic systems), then ${\bf B}^{\rm r}_k$
brings this past memory to the current assimilation cycle, improving the accuracy of
the resulting estimate of the true background-error covariance matrix ${\bf B}_k$.
Second, due to time smoothing (\ie averaging), the  sampling noise
is reduced,
leading to a more accurate estimate ${\bf B}^{\rm a}_k$.

\subsection {HHBEF: blending with static covariances}
\label{sec_hhbef1}

The first idea of the new hybrid-HBEF (HHBEF) 
filter  is to replace the HBEF's persistence forecast for the covariance
matrix ${\bf B}_{k}$ (Eq.(\ref{Bf})) with 
a regression-to-the-mean forecast:
\begin {equation}
\label{Bfc}
{\bf B}^{\rm f}_{k} =  w {\bf B}^{\rm a}_{k-1} + (1- w) {\bf B}^{\rm c}.
\end {equation}
Here, ${\bf B}^{\rm c}$ is the climatological  covariance matrix and 
$w \in [0,1]$ is the scalar weight.

\subsection {Spatial smoothing of the covariances}
\label{sec_hhbef2}

The second idea of HHBEF is to accommodate spatially smoothed ensemble covariances.
Here, we review the approach by
\citet{BuehnerCharron}, who studied spectral-space localization, found it useful in reducing
sampling noise, and noted that it is equivalent to a spatial smoothing of the covariance function ---
see their Eq.(9), which we rewrite here as
\begin {equation}
\label{sBsmo}
\check B(s_1,s_2) = \int_{\S^1(R)} \kappa(s) \,B(s_1-s, s_2-s) \d s,
\end {equation}
where  the check $\,\check{}\,$  designates the smoothing and  
$\kappa(s) \ge 0$ is the weighting (averaging) function.
In the space-discrete case, Eq.(\ref{sBsmo}) can be approximated  as follows:
\begin {equation}
\label{sBsmo2}
B_{ii'} = \sum_{s=-n/2+1}^{n/2} \kappa_s \,B_{i-s, i'-s},
\end {equation}
where $\kappa_s \ge 0$ are the weights such that $\sum \kappa_s=1$.
In the matrix form Eq.(\ref{sBsmo2}) can be written as
\begin {equation}
\label{msBsmo}
\check{\bf B}  =  \sum_{s=-n/2+1}^{n/2} \kappa_s \,{\cal F}^s {\bf B}  {\cal F}^{-s},
\end {equation}
where ${\cal F}$ is the forward-shift operator.
From  Eq.(\ref{msBsmo}) with non-negative $\kappa_s$ and given that 
${\cal F}^{-1} = {\cal F}^\top$,
it immediately follows that the smoothing 
is properly defined in the sense that the resulting
smoothed covariance matrix $\check{\bf B}$ is non-negatively definite.

If ${\bf B}=\frac{1}{N-1} \sum {\bf e}_l {\bf e}_l^\top$, 
where ${\bf e}_l$ is the $l$-th ensemble perturbation, $l=1,\dots,N$,
then Eq.(\ref{msBsmo}) implies that the spatial smoothing of the sample covariance matrix can be performed by
spatially shifting (through the repeated application of the forward and backward shift operators) and weighting 
(through the multiplication by $\sqrt{\kappa_s}$) ensemble perturbations --- as it  was proposed and tested by 
\citet{BuehnerCharron}.

\subsection {HHBEF's hybrid covariances}
\label{sec_hhbef3}

Combining Eq.(\ref{Bfc}) with Eq.(\ref{Ba}), where ${\bf B}^{\rm e}$ is replaced by its 
spatially smoothed version $\check{\bf B}^{\rm e}$ computed according to Eq.(\ref{msBsmo}), 
yields the HHBEF's analog of the HBEF's Eq.(\ref{tBsmo}):
\begin {equation}
\label{Bsmo_hhbef}
{\bf B}^{\rm a}_k = \mu w {\bf B}^{\rm a}_{k-1} +(1-\mu)\check{\bf B}^{\rm e}_k +
                     \mu (1-w) {\bf B}^{\rm c},
\end {equation}
where  $\mu$ is the weight of  ${\bf B}^{\rm f}_k$
relative to $\check{\bf B}^{\rm e}_k$.
Solving  Eq.(\ref{Bsmo_hhbef}) (which is a forced linear difference equation; the derivation is omitted)
shows that, for $\mu w <1$ and after an initial transient, ${\bf B}^{\rm a}_{k}$ 
has the four components:
\begin {equation}
\label{Baw}
{\bf B}^{\rm a}_{k} =  w_{\rm e} {\bf B}^{\rm e}_{k} + 
                       w_{\rm es}{\bf B}^{\rm es}_{k} + 
                       w_{\rm r} {\bf B}^{\rm r}_{k} +   
                       w_{\rm c} {\bf B}^{\rm c},
\end {equation}
where $w_{\rm e} = (1-\mu)\kappa_0$ 
is the weight of the current ensemble covariances ${\bf B}^{\rm e}_{k}$,
$w_{\rm es} = (1-\mu)(1-\kappa_0)$ 
is the weight of the {\em spatially smoothed} 
current ensemble covariances 
${\bf B}^{\rm es}_{k} =  \check{\bf B}^{\rm e}_{k} - {\bf B}^{\rm e}_{k}$,
$w_{\rm c} = {\mu (1- w)}/{(1 -\mu w)}$ 
is the weight of the climatological covariances ${\bf B}^{\rm c}$, 
and 
$w_{\rm r} = \mu w(1-\mu)/(1-\mu w)$ 
is the weight of the spatiotemporally-smoothed recent-past 
ensemble covariances ${\bf B}^{\rm r}_{k}$:
\begin {equation}
\label{Br}
{\bf B}^{\rm r}_k = \frac{ \check{\bf B}^{\rm e}_{k-1} + \mu w \check{\bf B}^{\rm e}_{k-2} + 
                                               (\mu w)^2 \check{\bf B}^{\rm e}_{k-3} + \dots  }
                         { 1 + \mu w + (\mu w)^2+\dots }.
\end {equation}
The smoothing time scale (measured 
in assimilation cycles) is $-1/\log(\mu w)$.

Equations (\ref{msBsmo}) and (\ref{Bsmo_hhbef}) show that    
$w=1$ and $\kappa_s=\delta_{s0}$ (where $\delta_{s0}$ is the Kronecker delta) reduces HHBEF to HBEF,
$w=0$ to EnVar,
$\mu=0$ and $\kappa_s=\delta_{s0}$ to pure EnKF,
and $w=0, \mu=1$ recovers the filter with static background-error covariances.

Thus, HHBEF employs blending of (localized) sample covariances 
with both climatological and spatiotemporally-smoothed background-error covariances.

\section {Performance of the three covariance-blending techniques  under non-stationarity}
\label{sec_perfHybr}

In this section, we experimentally study  how  temporal smoothing of background-error covariances, 
their spatial smoothing, and their mixing with climatology 
affects the performance of the classical stochastic EnKF \citep[e.g.,][]{Houtekamer2016review}
and HHBEF  under different
regimes of  non-stationarity.

\subsection {``Twin'' experimental methodology}
\label{sec_hybr_twin}

In the  experiments below, the truth was generated using the discretized 
Eq.(\ref{dsm1}), that is, 
Eq.(\ref{dxidt_dscr}):
$\boldsymbol\xi_k  = {\bf F}_k \, \boldsymbol\xi_{k-1} + \boldsymbol\varepsilon_k$,
where $\boldsymbol\varepsilon_k$ is the time and space-discrete model forcing.
The forecast operator ${\bf F}_k$ was available to the filters, so that, given the 
analysis $\boldsymbol\xi^{\rm a}_{k-1}$, the next-time-instant forecast was 
$\boldsymbol\xi^{\rm f}_{k} ={\bf F}_k  \boldsymbol\xi^{\rm a}_{k-1}$.
In this setting, the model forcing $\boldsymbol\varepsilon_k$ becomes the model error
(see, e.g., \citet{TsyGaySPG}), with its  covariance matrix ${\bf Q}_k$ also
available  to the filters.
We reiterate that in each experiment, the secondary fields (which determine the forecast
operator and the model-error statistics), once generated, were kept fixed.

The above setting implies that {\em non-identical} true and model twins were used,
with the only difference between the two models being the stochastic model error.
In contrast to using {\em different models}
to generate the truth and to perform the forecast, this approach
provides us with the exact knowledge 
(and in principle, full control) of 
the model error statistics.
It is this knowledge (along with  linearity) that justifies the use of the exact Kalman filter.

\subsection {Data assimilation setup}
\label{sec_DA_setup}

The default DSADM setup was the same as in section \ref{sec_mdl_setup}.
The default ensemble size was $N=10$.
The setup of the filters  and  the observation network
were chosen with the intention to mimic --- as much as possible with a 1D model ---
the setup of a realistic meteorological data assimilation scheme.

The  time interval between consecutive
analyses, $T_a$, was  selected using the following criterion.
The DSADM field correlation with lag $T_a$ should approach the typical 
meteorological-field correlation at lag 6h 
(the most widely used in operational practice assimilation cycle in global schemes).
On the one hand, the mid-tropospheric  6h
time correlation was estimated by \citet[][Fig.1(b)]{Seaman} 
using radiosonde data to be about 0.8 for the wind fields 
 and can be interpolated to the value of 
about 0.95 for geopotential  \citep[][]{Olevskaya} (that is, about 0.9 on average).
On the other hand, this 0.9  correlation in the  DSADM field (with the default setup) occurred,
on average, at the 12h lag.
So, we set $T_a=12$h. This is twice the typical assimilation cycle in real-world systems and 
is consistent with the selection of $\overline L$   and $\overline T$  
in section \ref{sec_mdl_setup} 
(about twice as large as the respective typical scales of meteorological fields).

Observations were generated every assimilation cycle 
at every 10th spatial grid point.
The observation error standard deviation (=6) was 
selected to ensure that the mean reduction of the forecast-error variance  in the analysis
was close to the value of 10\% reported by \citet[][section 8]{Errico2014} for a realistic
data assimilation system. 

The weights  of the spatial covariance smoothing $\kappa_s$ (see Eq.(\ref{msBsmo})) were specified to be a triangular function of 
the  spatial shift $s$, so that  $\kappa_s$ had a maximum at $s=0$ and 
vanished for $|s| > s_{\rm max}$.
Thus, the spatial smoothing was controlled by the single  parameter $s_{\rm max}$  
(the length scale of the spatial smoothing measured in grid spacings).

The covariance localization was performed using
the popular  function by \citet[][Eq.(4.10)]{Gaspari}.

\subsection {Setup of experiments}
\label{sec_expm_setup}

The relative roles of the three covariance-blending devices were studied by
measuring, first, the performance of EnKF  with each device switched on (in turn) \vs the pure EnKF,
and second, by measuring the performance of HHBEF with each 
device switched off (again, in turn) \vs the full HHBEF.
Switching on (off) the mixing with climatology is denoted by $+C$ ($-C$).
Similarly, switching on (off) the spatial smoothing of background-error covariances
is denoted by $+S$ ($-S$) and the temporal smoothing by $+T$ ($-T$).
Technically, all  filters were built within HHBEF.

Each filtering configuration was defined by the triple of the covariance-blending parameters 
$w$, $\mu$, and $s_{\rm max}$ as indicated in Table \ref{Tab_cnf}.
In each experiment, the free parameters marked in the respective row of Table \ref{Tab_cnf}
as ``Tuned'' as well as the multiplicative covariance inflation
and the length scale of the covariance localization 
were all simultaneously manually tuned to get the best performance.
The tuning was sometimes tedious but the optimum was always well defined and there 
was no indication that it was not unique.
The search for the optimum was facilitated by the observation 
that the better the performance of a filter, the less 
need for any covariance-regularization device.

\begin{table}[ht]
\caption{Configurations of  filters}

\begin{center}
\begin{tabular}{|l|c|c|c|}

\hline
Filter & $w$ & $\mu$ & $s_{\rm max}$\\
\hline
EnKF             & No effect       &  0        &  0  \\
EnKF $+C$ (EnVar)   & 0             &  Tuned &  0  \\ 
EnKF $+S$           & No effect     &  0        &  Tuned  \\
EnKF $+T$ (HBEF)    & 1  &  Tuned &  0  \\
\hline
HHBEF            & Tuned  &  Tuned    &  Tuned  \\
HHBEF $-C$          & 1       &  Tuned    &  Tuned  \\
HHBEF $-S$          & Tuned  &  Tuned   &  0  \\
HHBEF $-T$          &  0       &  Tuned   &  Tuned  \\
\hline
\end{tabular}
\end{center}
\label{Tab_cnf}
\end{table}

The performance of a filter, ${\rm f}$, was measured by its background root-mean-square error (brmse,
with respect to the known truth,
in runs with 5000 assimilation cycles) relative to that of the exact  Kalman 
filter (KF):
\begin {equation}
\label{perf}
{\rm rel.err}({\rm f}) =  \frac { {\rm brmse}({\rm f}) - {\rm brmse}({\rm KF}) } { {\rm brmse}({\rm KF}) }.
\end {equation}

The climatological background-error covariance matrix 
${\bf B}^{\rm c}$ was specified in each experiment 
to be the time-mean KF's background-error covariance matrix
(in a run with 50,000 assimilation cycles) --- for simplicity and in an attempt 
to reproduce the realistic situation
in which only a proxy to  ${\bf B}^{\rm c}$ is  available.

\subsection {Results}
\label{sec_hybr_strength}

As  noted in section \ref{sec_intext}, 
the non-stationarity grows with the increasing external parameters
$\sd (U^*)$, $\varkappa_\rho, \varkappa_\nu,\varkappa_\sigma, \pi_\rho, \pi_\nu$.
Here we compare the filters in  four regimes with different strengths of non-stationarity
as detailed in Table \ref{Tab_strength}.

\begin{table}[ht]
\caption{External parameters of DSADM in the four regimes of non-stationarity}

\begin{center}
\begin{tabular}{|l|l|c|c|c|c|c|c|}

\hline
\# & Regime & $\sd (U^*)$ & $\varkappa_\rho=\varkappa_\nu=\varkappa_\sigma$ & $\pi_\rho$ & $\pi_\nu$\\
\hline
0 & Stationary                & 0   & 1  &  0    &  0  \\
1 & Weakly non-stationary     &  5  & 2  & 0.01  & 0 \\
2 & Default non-stationary    & 10  & 3  & 0.02  & 0.01  \\
3 & Strongly non-stationary   &  20 &  6 & 0.04  & 0.02 \\
\hline
\end{tabular}
\end{center}
\label{Tab_strength}
\end{table}

\begin{figure}
\begin{center}
   { 
        \scalebox{0.5}{ \includegraphics{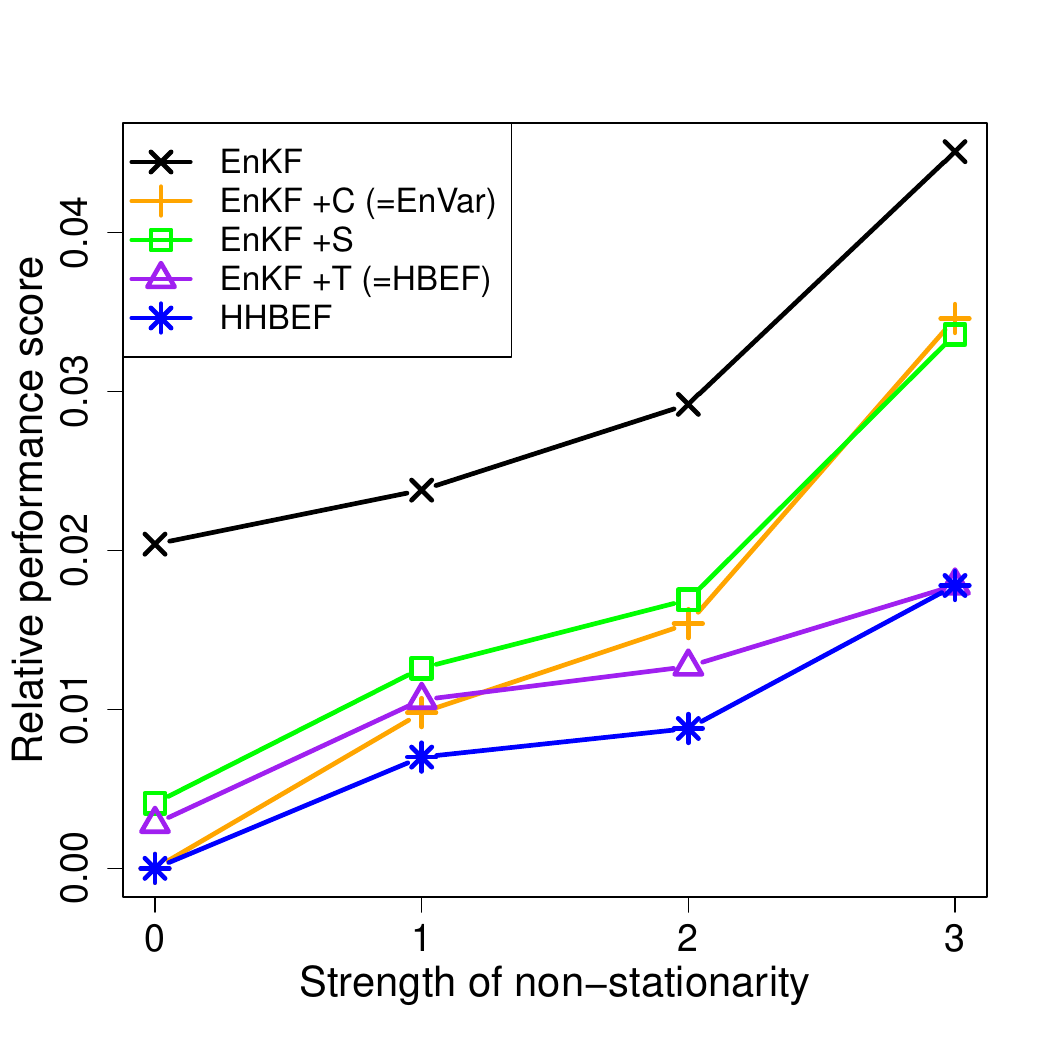}}
    }
\end{center}
  \caption{Performance scores (relative to KF, the lower the better) of the pure EnKF (the upper curve),
   EnKF with one of the three covariance-blending devices switched on (the three lower curves),
   and HHBEF (the lowest curve).
  The filters' notation follows Table \ref{Tab_cnf}.
  The numbers on the x-axis correspond to the numbers in the first column of Table \ref{Tab_strength}.
              }
\label{Fig_Eplus_strength}
\end{figure}

Figure \ref{Fig_Eplus_strength} shows the performance scores (defined in Eq.(\ref{perf})) 
for the four configurations indicated in the upper half of Table \ref{Tab_cnf} plus the full HHBEF.
One can see that all three covariance-blending techniques were quite successful in improving  the performance of EnKF. This could be expected because the ensemble size $N=10$ was rather small.

Static covariances were most useful in the low non-stationarity regimes 
(non-stationarity strengths 0 and 1) while being least useful under strong non-stationarity. 
This is reasonable because under stationarity,
static covariances suffice for exact filtering, whilst they become useless 
when the filtering-error statistics are, typically, very different from climatology,
which is the case
in a highly non-stationary regime.

The temporal covariance smoothing is seen to be  beneficial in all regimes, 
 especially under strong non-stationarity, and always more efficient than the
spatial covariance smoothing.
The reason for the observed systematic advantage of the temporal covariance smoothing 
over the spatial covariance smoothing 
is unclear.
We may conjecture that neighboring-in-space covariances bring less information because they are
part of the same sample covariance matrix, whereas neighboring-in-time covariances come from 
a different sample covariance matrix and  can therefore be more independent.

It is seen in Fig.(\ref{Fig_Eplus_strength}) that HHBEF was more accurate than the other filters.

\begin{figure}
\begin{center}
   { 
       \scalebox{0.5}{ \includegraphics{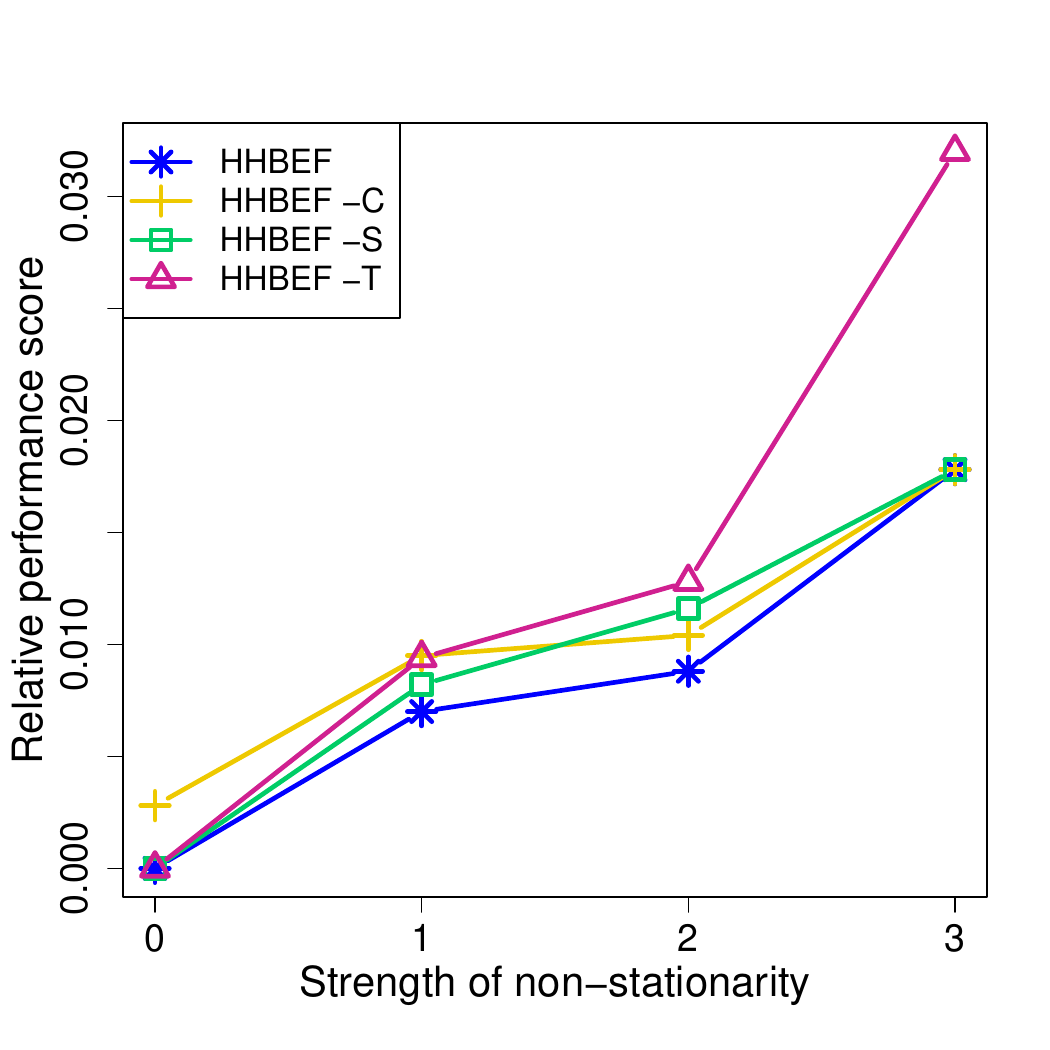}}
    }
\end{center}
  \caption{Same as Fig.\ref{Fig_Eplus_strength}, but for the  full HHBEF and HHBEF with one 
  of the three covariance-blending devices switched off.
              }
\label{Fig_Hminus_strength}
\end{figure}

To check if the three covariance-blending techniques remained beneficial
when used in combination, the four configurations indicated in the lower half of Table \ref{Tab_cnf} 
were compared: HHBEF \vs HHBEF with one of the three covariance-blending devices withheld. 
Figure \ref{Fig_Hminus_strength} displays the results, which were, generally, consistent
with those depicted in Fig.\ref{Fig_Eplus_strength}.
Under non-stationarity, all  three blending devices remained  usefulr  if used jointly except for the strongly
non-stationary regime 3, where only time smoothing was beneficial.
Not using static covariances led to the worst performance under stationarity and low non-stationarity.
Switching off  the time smoothing was most detrimental for the default and  strong non-stationarity.

\begin{figure}
\begin{center}
   { 
       \scalebox{0.5}{ \includegraphics{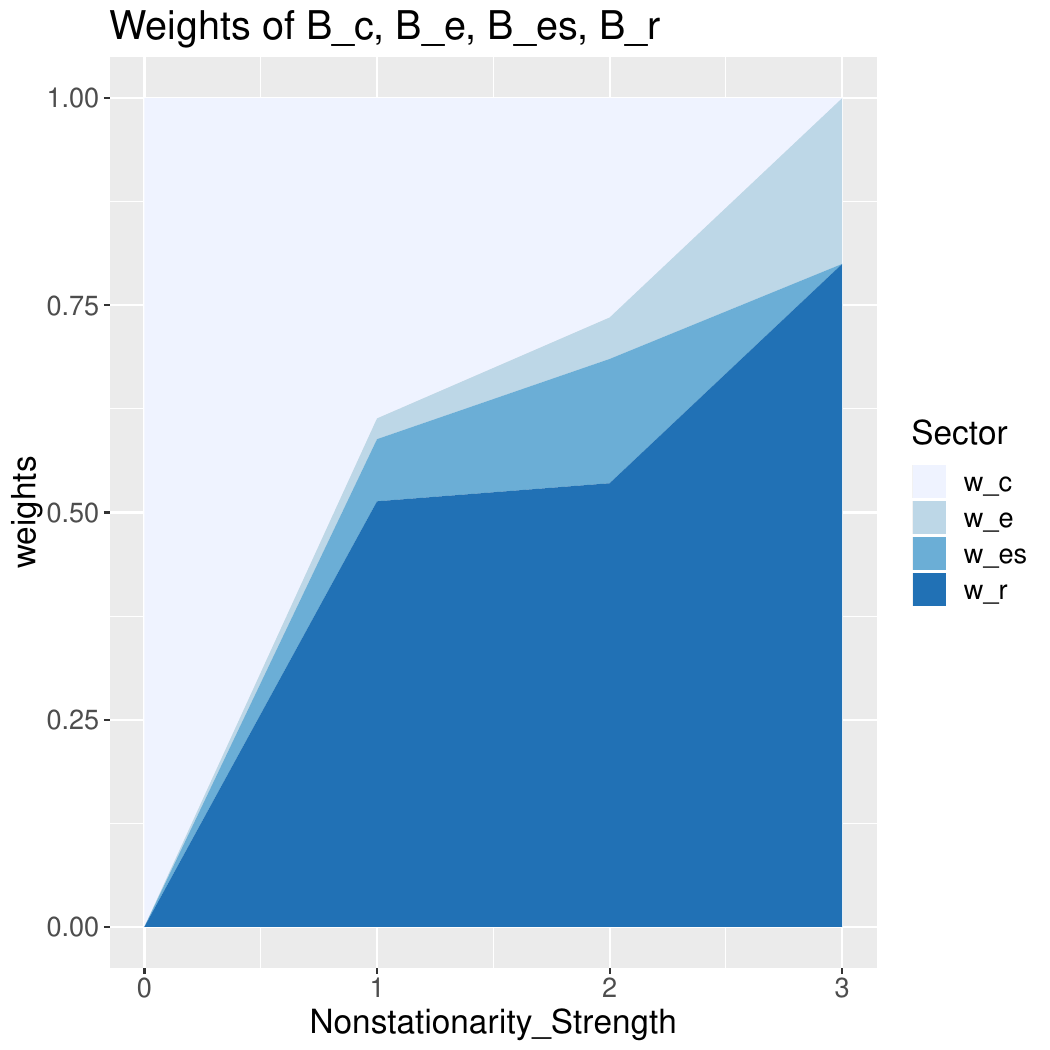}}
    }
\end{center}
  \caption{Optimal blending weights of climatological covariances,  $w_{\rm c}$,
  non-smoothed current ensemble covariances, $w_{\rm e}$,
  spatially smoothed  current ensemble covariances, $w_{\rm es}$, 
  and spatiotemporally-smoothed recent-past covariances, $w_{\rm r}$.
  The numbers on the x-axis correspond to the numbers in the first column of Table \ref{Tab_strength}.
              }
\label{Fig_hybr_weights}
\end{figure}

It is also of interest to inspect the optimal HHBEF's weights:   
 $w_{\rm e}$,  
$w_{\rm es}$,  
 $w_{\rm r}$, and 
 $w_{\rm c}$ (see Eq.(\ref{Baw}) and the paragraph after it).
The weights are plotted in Fig.\ref{Fig_hybr_weights} for the same four strengths of non-stationarity
as above.
Overall, we see that the two dominant sources of hybrid covariances were
static covariances (prevailing in the low non-stationarity regimes) and spatiotemporally-smoothed covariances 
(most important if  the non-stationarity was high).
The role of non-smoothed ensemble covariances increased with the growing non-stationarity.
This is meaningful because any covariance blending reduces sampling noise but distorts
the flow-dependent  signal carried by the non-smoothed ensemble covariances,
and this distortion grows when the covariances become more variable.
The spatially averaged covariances were beneficial only for non-stationarity strengths 1 and 2 
and useless under the strong non-stationarity (and of course under stationarity).
This is consistent with Fig.\ref{Fig_Hminus_strength}.

We  also studied how the effects of the three covariance-blending techniques 
depend on the {\em time and length scales} of  non-stationarity 
(determined by the external parameter $L^*$, see Table \ref{Tab_params}).
We found that the effects were quite stable, while
blending with spatially and spatiotemporally-smoothed covariances 
was more beneficial for larger $L^*$  (not shown).
This can be expected because space  and time smoothing
is worthwhile only if the  covariances vary smoothly in space and time.

Finally, we explored the performance of the three covariance-blending techniques for large ensembles
($N=80$ and $N=300$) under the default non-stationarity.
First, we found that for those ensembles, 
some covariance regularization was still useful.
This is reasonable because for the non-regularized sample covariance matrix to be a good estimator of the 
true covariance matrix, the sample size must be significantly greater than 
the dimensionality of the matrix,  see e.g., \citet[][Section 2]{BaiShi}, that is, $N\gg n$,
where, we recall, $n=60$ in our experiments.
Second, the effects of all five regularization techniques (the three blending devices plus
covariance localization and multiplicative covariance inflation) were significantly reduced as compared with the ensemble
sizes $N=10$ (see above) and $N=30$   (not shown), which is, of course, meaningful. 
Third, the effect of  blending with climatology was negligibly small
(because two factors ---  variability in the covariances and large ensemble size --- 
both reduce the usefulness of static covariances).
Fourth, the effects of  time and space smoothing  remained positive, with  
time smoothing being still more useful.
Fifth, only with huge ensembles ($N=3000$), EnKF needed no covariance regularization anymore  (not shown).

\section {Conclusions}

In this paper we have presented a new doubly stochastic
advection-diffusion-decay model (DSADM) on the circle.
Double stochasticity means that not only the model forcing is stochastic,
the model coefficients (parameters) are random as well.
The parameters are specified to be transformed Gaussian 
random fields, with each Gaussian field satisfying its own 
 stochastic advection-diffusion-decay model with constant coefficients.
Thus,  DSADM is hierarchical, built of linear  
stochastic partial differential equations at two levels in the hierarchy.
DSADM is designed to be used as a toy model of truth to
test and develop data assimilation methodologies.

The main advantage of  DSADM is its capability of 
generating spatiotemporal  random fields 
with tunable non-stationarity in space and time, while maintaining linearity and
Gaussianity. This allows one, first, to separate effects of  non-stationarity
from effects of nonlinearity and non-Gaussianity (which is impossible with nonlinear models of truth).
Second,  linearity and Gaussianity allow the use of the Kalman filter as an unbeatable benchmark,
which, again, is rarely possible with nonlinear models.
With  a small modification, DSADM can also be used to study the role of non-Gaussianity not caused by nonlinearity
(say, non-Gaussianity of observation or model errors).

We have used  DSADM to study the impact of  non-stationarity on the performance 
of the following three covariance regularization techniques in EnKF:  blending  with 
static, time-smoothed, and space-smoothed background-error  covariances.
DSADM and the synthetic observation network were set up  to resemble 
(as far as possible with a one-dimensional model)
a realistic meteorological data assimilation system.
We found that blending with static covariances was most beneficial in filtering
regimes with low non-stationarity, while  time-smoothing was most useful under
medium and high non-stationarity. 
Space-smoothing was  less efficient than time-smoothing. 
These findings were valid in a wide range of ensemble sizes.
The role of time and space smoothing was found larger when the length and time scales of  non-stationarity
patterns were larger than the respective scales of the background-error field itself.
A new filter termed HHBEF (Hybrid Hierarchical Bayes Ensemble Filter) that combines all three 
covariance-blending techniques proved to be  most accurate among all configurations of the filters tested.

We believe that these results are relevant for real-world applications, but of course
they are model dependent and the degree of  relevance remains to be verified,
especially because the model is new.

Thus, the optimal blend of ensemble covariances with climatology as well as with 
time-smoothed  and space-smoothed background-error covariances is found to strongly depend on 
characteristics of  non-stationarity.
How large is the actual non-stationarity of the spatiotemporal background-error field in practical data assimilation systems,
how large are the time and space scales of  non-stationarity patterns as compared to 
the respective scales of the background-error field itself, how non-stationarity
depends on the weather situation, season, scale, altitude,
meteorological field, observation density and accuracy---all
these questions remain open.
Addressing these questions may help optimize hybrid filters that 
accommodate various covariance regularization techniques.

\section* {Acknowledgments}

Valuable comments of two anonymous reviewers helped improve the paper 
significantly.





\bibliography{mybibfile}


\end{document}